\documentclass[10pt,a4paper]{article}
\usepackage{graphpap,epsfig,amssymb,graphicx,textcomp}
\usepackage[utf8]{inputenc}
\usepackage[english]{babel}
\begin{document}
\begin{center}
\large {\bf Cellular 
Automata Model of Synchronization in Coupled Oscillators}
\end{center}

\vskip 1cm

\begin{center}{\it Amitava Banerjee and Muktish Acharyya}\\
{\it Department of Physics, Presidency University,}\\
{\it 86/1 College street, Kolkata-700073, INDIA}\\
{E-mail: amitava8196@gmail.com}\\
{E-mail: muktish.physics@presiuniv.ac.in}\end{center}

\begin{abstract}
We have developed a simple cellular automata model for nonlinearly coupled phase oscillators which can exhibit many important collective dynamical states found in other synchronizing systems. The state of our system is specified by a set of integers chosen from a finite set and defined on a lattice with periodic boundary conditions. The integers undergo coupled dynamics over discrete time steps. Depending on the values of coupling strength and range of coupling, we observed interesting collective dynamical phases namely: asynchronous, where all the integers oscillate incoherently; synchronized, where all
integers oscillate coherently and also other states of intermediate and time-dependent ordering. We have adapted conventional order parameters used in coupled oscillator systems to measure the amount of synchrony in our system. We have plotted phase diagrams of these order parameters in the
plane of strength of coupling and the radius of coupling. The phase diagrams reveal interesting properties about the nature of the synchronizing transition. 
There are partially ordered states, where there are synchronized clusters which are shown to have a power law distribution of their sizes. The power law exponent is found to be independent of the system parameters. We also discuss the possibility of chimera states in this model. A criterion of persistence of chimera is developed analytically and compared with numerical simulation.
\end{abstract} 

\vskip 1 cm

\vskip 1 cm

{\bf I. Introduction:}

The identification, classification and characterization of 
collective dynamical states in a population of nonlinearly 
coupled oscillators is an interesting and old problem\cite{fermi,strogatz1,waller}. For many of such systems, we are concerned about only the phases of the individual oscillators, especially if the system is composed of limit cycle oscillators. In that case, the phase space dynamics of a single oscillator takes place mainly near a limit cycle and so
 the amplitude of oscillation can be considered to be practically constant during the time evolution. If we couple a number of such oscillators, then, under certain coupling conditions, these systems may show complete phase-synchronized state. Starting from a random distribution of phases, when the total system synchronizes, its individuals oscillate in time in such a way that their phases are almost equal or distributed over a very narrow range\cite{strogatz2,rosin,bonilla}. 
Apart from being of mathematical interest, these systems are often used 
to model many natural systems which admit spontaneous synchronization by the interaction of their individuals. Some notable examples include collective dynamics of pulse-coupled excitable neurons\cite{masuda,mirollo}, cardiac dysrhythmias\cite{supru,Guevara} and spontaneous synchronization of the flashes of fireflies\cite{strogatz2,strogatz3} among many. 

A special class of such problems, the Kuramoto model 
for coupled phase oscillators is an well-studied problem in nonlinear dynamics\cite{kuramoto1}.
 The model consists of $N$ coupled phase oscillators whose 
phases $\theta_{i}$ obey the set of $N$ first-order ordinary 
differential equation 

$$\dot{\theta_{i}}= \omega_{i}+\Sigma_{j=1}^{N}K_{ij}sin(\theta_{j}-\theta_{i}) \eqno(1.1)$$

with $i=1,2,..,N$ where $\omega_{i}$ is the natural frequency of oscillation 
of the $i$-th oscillator and the matrix $K$ denotes the coupling between 
the oscillators. The model is widely used as a paradigm to study phase 
synchronization of oscillators\cite{strogatz4,acebron}. Previously, it was generally believed that systems of identical oscillators, in which the coupling is also identical for all oscillators exhibits global phase synchronization as the only possible collective state. However, recently it was discovered that 
for identical oscillators and for certain types of non-local coupling between 
them, the system self-organizes to a stable coexistence of phase-synchronous and 
asynchronous groups of oscillators, a state known by the name `chimera state'\cite{kuramoto2}.
 Subsequently, this state was found in many other systems, described by 
differential equations as well as coupled map lattices\cite{panaggio,omelchenko1,hagerstrom}. At present, the stability, bifurcation and spatio-temporal dynamics of chimera states are being studied both numerically\cite{oleh,seiber,zakharova} and analytically\cite{abrams1,abrams2}. Apart from these studies, chimera states are also demonstrated experimentally in chemical oscillators\cite{tinsley,nkomo,panaggio} and 
hypothesized to be related to the unihemispheric sleep found in some species of 
birds and mammals\cite{ma,panaggio}, power grid failures due to partial or complete loss of synchronization of power generators\cite{panaggio} or in social systems where consensus is formed in a certain fraction of the population, while the other fraction remains disordered in following certain trends or making decisions\cite{panaggio,gonzalez}.  

In this work, we describe synchronous as well as partially synchronous stable states found 
in a very simple {\it cellular automata} \cite{wolfram} model of phase oscillators. It is only recently that a work on chimera states in a 
cellular automata model is published \cite{vlad1}, but the model used in that work is of an altogether different kind from ours, so we do not go to
the details. Inspired by that work, we propose our model which also shows many interesting collective dynamical states. In our model, we assume a linear 
lattice of oscillators with their phases represented by their integral automata values. 
The automata rule gives the dynamics of the phases in discrete time steps and we use a minimally 
nonlinear coupling given by the modulus of the phase difference of a pair 
of oscillators. Starting from random automata values, the model is shown 
to produce synchronized states as well as partially ordered states 
similar to those found in systems governed by differential equations, as we 
vary the coupling radius. Furthermore, the transition from asynchronous 
to phase-synchronized state is clearly characterized by a sudden increase 
of the order parameter resembling the corresponding transition occurring 
in systems governed by differential equations. A statistical analysis of 
synchronized clusters in the partially synchronized state reveals that 
the cluster size distribution follows a power law, with an exponent which is independent of the parameters of the system.  

This paper is organized as follows: section II describes the automata 
formalism used to model coupled phase oscillators, compares it to 
conventional models and describes the results of the simulation, 
especially the statistical characterizations of various collective 
dynamical states and section III discusses about conclusions and various possible future extensions of the model.

\vskip 1cm

\noindent {\bf II. Cellular Automata Model and Numerical Simulation Results:}
In this paper, we have tried to construct a cellular automata model representing a system of coupled phase oscillators. In doing so, we seek a model which will be as simple as possible in its description and automata rules so as to minimize computational cost, but yet will capture all of the properties and collective behaviour seen in those systems. Our proposed model is as follows:
we have considered a linear array of $N$ number of 
coupled phase oscillators under periodic
boundary condition having identical natural frequency. The 'phase' of i-th oscillator at any instant ($t$) is being
represented by automaton $P_i(t)$, which can take values from a finite set of positive integers only, or be equal to $0$. To maintain the periodicity in the phase variable, all $P_i$'s assume values between
$0$ and a fixed integer, say $l$ and are defined modulo $l$. This is essential to connect our model to those of phase oscillators. In our system, the oscillators are coupled to their left and right neighbours symmetrically and by homogeneous and time-independent coupling. The number of such coupled oscillators 
at any side (left/right) of one oscillator is, say $N_c$, with the ratio $r=N_c/N$ $(0\leq r\leq 0.5)$ being called the radius of coupling.
Initially (at $t=0$), the automaton values of all the oscillators are chosen randomly
(uniformly distributed) between $0$ and $l$, such that, the dynamics of 
coupled oscillators starts with totally asynchronous state.
The time evolution of phase of i-th oscillator is
represented by following automata rule:

\begin{equation}
P_i(t+1) = P_i(t) + \Big\lfloor {{\epsilon} \over {2N_c}}\Sigma_{j=i-N_c}^{i+N_c}|P_i(t)-P_j(t)|\Big\rfloor 
\end{equation}
where the floor function $f(x)=\lfloor x\rfloor $ implies the greatest integer not exceeding $x$
and for any $P_i(t+1) > 50$, 

\begin{equation}
P_i(t+1) = {\rm Mod} [P_i(t+1),50].
\end{equation}

\noindent Here, $\epsilon$ is the strength of coupling and it is assumed to be positive and independent of oscillator index. Here we have updated 
the phases (automaton $P_i$) by using parallel updating scheme. The automata rule equation is written from the rotating frame moving with the common natural frequency of the oscillators, hence the constant natural frequency term is absent in it. It may be noted that the coupling is of nonlinear type, but perhaps of the simplest form. It is truly remarkable that the system with such a simple nonlinearity can exhibit such a rich spectrum of dynamical behaviour.

In our simulation, we have first considered $N=100$ number of oscillators and assumed $l=50$. Depending
on the values of coupling strength ($\epsilon$) and coupling radius ($r$), we
have observed various dynamical states of the oscillators. For small,
values of $\epsilon$ and $r$, the oscillators are seen to oscillate 
incoherently without any particular relation to each other, showing an asynchronous dynamical state even after a 
long time ($t=40000$). A snapshot is shown in Fig-1(a). 
A typical video may be found
in ref. \cite{video1}. For sufficiently large values of 
$\epsilon$ and $r$, we have found a complete 
synchronous state, when all the oscillators are found to 
oscillate coherently. 
A typical video may be found in ref. \cite{video2}. Fig-1(c)
shows snapshot of such synchronized state of the oscillators. Other interesting states
were observed in the intermediate range of values of $\epsilon$ and $r$. Here,
for a range of $\epsilon$ and $r$ values, we observed that multiple groups of
connected oscillators oscillate in a synchronized fashion. A snapshot is in Fig-1(b) and the dynamical evolution of 
this state can be found in a video in ref. \cite{video3}. The snapshot of the state indeed looks like a multichimera state.

These dynamical states are observed even at larger systems with $N=1000$ at similar values of the parameters $\epsilon$ and $r$. We also produce the simulation results for these cases as space-time plots of phases $P_i$ and nearest neighbour phase differences $P_i-P_{i+1}$ (calculated using the periodic boundary condition at the right end of the lattice) , together with snapshots at some particular time in Figs-2,3,4. Fig-5 shows the initial random distribution of phases from which these self-organized.

Interestingly, in both the cases for $N=100$ and $N=1000$, we see that the partially ordered states (like the asynchronous or the synchronized states) are stable in time, in the sense that they do not evolve to synchronized or asynchronous states, at least for a very large time. However, as is evident from the figure 4, the fraction of the oscillators locked in a synchronized cluster varies in time. At times, the system closely resembles a completely incoherent state till the synchronized clusters turn up again. Also, the synchronized clusters may contain different individuals at different times. These properties are analogous to the corresponding behaviour of `breathing chimeras', which are observed earlier in Lorenz systems \cite{gopal1} as well as in Kuramoto Systems \cite{panaggio} or laminar-turbulent transitions. A direct visual identification of this state as a multichimera may not be possible, and, perhaps this comparison should not be expected too strictly as this is a new system altogether. The correspondence, however, becomes more closer in terms of certain statistical quantities developed to distinguish between various states in continuous systems. We shall discuss about them shortly. 

In our model, the well-known single-cluster chimera states seen in many coupled oscillators systems, however, is hard to achieve by self-organization from a random initial configuration. Rather, we have carried out a study regarding the stability of chimera states at various parameter values, which reveals why these states are so elusive in this model. The following analytic study will inspire future directions of the modifications of this model to other models involving chimera states. We first try to estimate analytically what is the condition for a synchronized cluster in a chimera state to be static, i.e., if the $k$-th oscillator belongs to this cluster at some time, say, $t=0$, then, $P_k(t+1)=P_k(t)$ for all subsequent times $t$. Because we wrote equation $1$ from a frame moving uniformly in time with the frequency equal to the natural frequency of the oscillators, a static cluster possesses only this motion. We note that while this case is easy to handle, analytically, chimera states can also exist if this condition is not satisfied, but the $P_k$'s evolve over time in such a manner that the difference between their values for neighbouring oscillators remain zero for all times and consequently, the cluster remains synchronized at later times. Nonetheless, the former condition yields a relation between number of phase levels $l$ and the parameter $\epsilon$, which is expected to guide us to the parameter regime where we may discover some interesting features.

We start by a simplifying assumption that a chimera state is a coexistence of a synchronized cluster of size $n$ with the rest of the $N-n$ oscillators having their automaton values $P_i$ uniformly distributed in the range $[0,l]$ at all times. In that case, if the $k$-th oscillator belongs to the cluster, then for it, the condition $P_k(t+1)=P_k(t)$, using equation $1$ and the definition of the function $f(x)=\lfloor x \rfloor$, becomes
\begin{equation}
{{\epsilon} \over {2N_c}}\Sigma_{j=k-N_c}^{k+N_c}|P_k(t)-P_j(t)| < 1.
\end{equation}

Based on this expression, we find a simple criterion for the cluster to be stable. In order to estimate the value of the quantity in the left-hand-side of the inequality, we assume that in our simplified model of chimera state, all
the oscillators with index $1$ to $n$ are locked in the synchronized cluster and are fixed at a particular phase level $c (0\le c \ge l)$, whereas the rest are 
uniformly distributed between $0$ and $l$. We also assume, for simplicity, that $N_c<n$. 

In this case, we find, for the oscillators belonging to the cluster, the only contributions to the sum 

$${{\epsilon} \over {2N_c}}\Sigma_{j=k-N_c}^{k+N_c}|P_k(t)-P_j(t)|$$

 comes from the terms corresponding to the oscillators with index $j$ which are outside the cluster. This is because
all the $P_i$'s have the same value in the cluster, and so their phase difference is zero. Hence, as $N_c<n$, so the maximum value for this term comes for the oscillator which lies at the side of the chimera and is coupled to $N_c$ oscillators whose phases are uniformly distributed. If this maximum value is $<1$, then so are other values. So a sufficient condition for the cluster to be static is to have the maximum value of the sum to be less than $1$. For a large system, we replace the above summation by the number of nonzero terms in the summation, multiplied
by the expectation value of each term,i.e.,
\begin{equation}
{{\epsilon} \over {2N_c}}\Sigma_{j=k-N_c}^{k+N_c}|P_k(t)-P_j(t)| = {{\epsilon} \over {2N_c}} \times N_c \times \left <|c-x|\right >
\end{equation}
where $x$ is an integral random number taking values uniformly from the range $[0,l]$. Calculating the expectation over the uniform distribution, we rewrite the above inequality (3) as
\begin{equation}
\frac{\epsilon}{4(l+1)}[l^2+l+2c^2-2lc]<1. 
\end{equation}
Interestingly, this result does depend neither on the coupling radius, nor the system size. However, these terms will be included in more sophisticated calculations. To check our limit, we use numerical simulations of equation $1$.
As one typical case, we take $N=1000$, $N_c=149$,$n=700$, $l=50$ and $c=26$. These correspond
to the inequality $\epsilon<0.15$ as a sufficient criterion for the cluster to be static, provided the rest of the oscillators remain
uniformly distributed for all later times. Indeed, from simulations (Fig-6), we see that the chimera
is almost static for $\epsilon<0.19242$ (which is very near our estimated value, which was a sufficient but not a necessary limit under crude assumptions) and distinguishable till the rest of the oscillators synchronizes (in contrary to our assumptions for the analysis), and for $\epsilon$ only slightly beyond this limit, the chimera is slowly destroyed and a multichimera-like state is formed. This delicate limit on
$\epsilon$ shows how fragile the chimera state is -- the single synchronized cluster either loses its identity into the globally synchronized state or is taken over by the asynchronous mass. Perhaps because of this phenomena, this state is hard to find directly from a random configuration. This behaviour has been checked for a range of values of $r$ and $N$. The exact value of $\epsilon$ at which this transition happens depends on system size, coupling radius, initial conditions etc.; but the sharp demarcation exists always. In this regard, we mention that the conventional chimera states found in other systems often also have a finite lifetime (which is a function of the finite system size) and lead to a collapse to a globally synchronous state \cite{col1,col2}. 

In the next parts, we shall discuss various statistical measures we have used to identify and distinguish between various states.
At first we describe the statistical measures to distinguish and identify the synchronous state.                   
Following the previous works\cite{strogatz4,acebron}, the synchronizing transition can be characterized by a complex order parameter $R$ given by the equation
\begin{equation}
Re^{i\psi }=\frac{1}{N}\Sigma _{j=1}^{N}e^{iP _{j}}
\end{equation}
 where $\psi$ is the average phase of the oscillators. It is to be noted here that even if our phase levels are discrete and defined to be in the interval ${(0,l)}$ unlike the natural range ${(0,2\pi)}$ 
of phases, the order parameter is still expected to be $0$ at desynchronized state and $1$ at synchronized state. We also define another 
order parameter $R'$, namely the ratio of the standard deviation of phases to the mean phase (normalized standard deviation) 
which is expected to be zero at the synchronized state. This is given by the equation

\begin{equation}
R'=\frac{\sqrt{\frac{(\Sigma _{j=1}^{N}P_j^2}{N})-(\frac{\Sigma _{j=1}^{N}P_j}{N})^2}}{\frac{\Sigma _{j=1}^{N}P_j}{N}}
\end{equation}

Both of these quantities measure global coherence of the system and neglects any local order. We use both quantities to investigate the onset of synchronization with varying coupling radius $r$. The occurrence of synchronization is sudden, as shown by the sharp rise of the squared argument of the order 
parameter and sharp dip of the normalized standard deviation roughly at a same coupling radius in Fig-7. These behaviour of the 
order parameter resembles those reported in previous synchronizing system governed by differential equations, e.g. the explosive synchronization in the 
Kuramoto model.\cite{strogatz4,acebron}

We now proceed to characterize the partially ordered dynamical states by parameters used in previous works\cite{gopal1,gopal2}. Here, we have used $N=200$ oscillators. We have used the
parameter $S$, the strength of incoherence, defined as follows\cite{gopal1,gopal2}: At any time step,
we have considered the new variable $Z_i=P_i-P_{i+1}$ representing the deviation of the phase of the $i-$th oscillator from its neighbor to the right. Then we divide the $N$ oscillators in $N_g$ number of groups, so that
each group contains $N_b=N/N_g$ number of oscillators. Next we calculate, the 
variance of $Z_i$ in k-th group as follows:

\begin{equation}
\sigma_k=  {\sqrt {{{1} \over {N_b}}{\sum_{i=1+(k-1)N_b}^{kN_b}
(Z_i-{\bar Z^k})^2}}}
\end{equation}

where we have defined the group average of $Z_i$ in the $k$-th group as

\begin{equation}
 \bar Z^k=\frac{\sum_{i=1+(k-1)N_b}^{kN_b}Z_i}{N_b}.
 \end{equation}

\noindent A new variable in each such group is introduced as
\begin{equation}
\lambda_k=\Theta (\delta - \sigma_k)
\end{equation}

\noindent where $\delta$ is a sufficiently small value and $\Theta$ stands for Heavyside step
function. With this, the strength of incoherence $S$ is defined as

\begin{equation}
S = 1 - {{\Sigma_{k=1}^{k=N_g} \lambda_k} \over {N_g}}.
\end{equation}

\noindent Here, in previous coupled oscillator models $S=1$ represents complete incoherence, $S=0$ represents
complete coherence and $0<S<1$ represents chimera or multichimera states. 

For further distinction between single chimera and multichimera states
(where $0<S<1$) one employs the discontinuity measure defined as\cite{gopal1}

\begin{equation}
\eta = \frac{(\Sigma_{k=1}^{k=N_g} |\lambda_k - \lambda_{k+1}|)}{2}
\end{equation}

\noindent with $\lambda_{N_{g}+1}=\lambda_{1}$ due to applied periodic boundary
condition. Here, $\eta=1$ for single chimera state and positive integer
between $1$ and $\frac{N_{g}}{2}$ for multichimera state. While our system does not show prominent chimera and multichimera states, the two measures indeed probes local ordering and spatial non-homogeneity in the system and thus gives a lot of information. For convenience, we show some examples of how the typical characteristic dynamical states in our system can be differentiated by these parameters in Fig-8.

By using these measures, we have shown the phase plots generated by the values of $S$ for varying
$\epsilon$ and $r$ after a large number of iterations in Fig-9(a) where
the regions of synchronized, asynchronous and partially ordered states
are identified. We have also distinguished the
states by the discontinuity measure plot and shown in Fig-9(b). In both plots, the sharp transition to synchronization with varying coupling radius is evident and it is also seen to be independent of the coupling strength $\epsilon$ for sufficiently large values of $\epsilon$. Thus the behaviour of the order parameter as seen in Fig-7 is generic, for a large range of values of $\epsilon$. Furthermore, the plots reveal that as the system undergoes the synchronizing transition from the asynchronous state with increasing coupling radius, at first, 
most of the systems form chimera-like states, which show smaller number of synchronized clusters in the system and hence relatively lower amounts of spatial discontinuity in the automata values. With increasing $r$, the initial clusters fragment into more clusters and the system's higher spatial discontinuity yields higher values of the discontinuity measure. Finally, then global synchronization sets in after a critical coupling radius. This sequence is perhaps not very counter-intuitive, especially if one considers the fact that the multi-cluster states have less broken global translational symmetry from the synchronized states when compared to chimera states.
The phase diagrams in Fig-9 also show that the system is unable to attain synchronization even for global coupling if the strength of coupling $\epsilon$ is lower than some critical value. However, when it attains synchronization, the critical coupling range is almost independent of the coupling strength. 
                                             
We previously remarked that the partially ordered states we have obtained is of a breathing nature. We justify this fact quantitatively by the observation that in these states, the value of strength of incoherence hugely varies in time as shown in Fig-10.

At this point, it becomes important to describe one particular feature of the dynamics. We have observed that it is multistable for the almost entire range of parameters. This multistability is expressed in the observation that depending on initial conditions, for a fixed value of the parameters, the trajectories in the system converge to different configurations asymptotically. The difference in those final configurations may be macroscopic, i.e., reflected in the order parameter or other statistical quantities like Discontinuity Measure or Strength of Incoherence; or they can be statistically indistinguishable by these quantities, differing only in the microscopic details involving the automaton values of the individual oscillators (for example, for two initial configurations, the system synchronizes in both the cases, but the integer equal to the final automaton value of the synchronized oscillators differ). As in complex systems like this one, one is generally interested about a statistical description of dynamical states, the former kind of multistability is more important. We proceed to study it further and simulate equation $1$ with $N=100$ starting from various different random initial conditions.

 We discover that the globally synchronous states, as well as the completely asynchronous states are the most common dynamical states at the values of the parameters for their occurrence suggested by the phase plots. In particular, in the parameter range $\epsilon \in [0.6,1.0]$ and $r\in [0.46,0.49]$ corresponding to region of synchrony, we chose $1000$ random initial configurations and out of them, only $37$ configurations failed to give the values $S=0,\eta=0$ (which represent synchronization) after $15000$ steps. As we noticed, the time to synchronize depends on initial configuration and parameter values (though not in a very systematic way), so it is not clear whether these small fraction of configurations can take longer time to synchronize or do not synchronize at all. Finding the distribution of the time to synchronize, as well as studying how that depends on the parameters would be interesting.
 
 On the part of the phase diagram showing asynchronous state,  in the parameter range $\epsilon \in [0.2,1.0]$ and $r\in [0.01,0.1]$ corresponding to region of asynchrony, we chose $4500$ random initial configurations and out of them, only $10$ configurations failed to give the values $S=1,\eta=0$ (which represent synchronization) after $15000$ steps. 
 
 The last two paragraphs show that synchronization as well as incoherence are very robust to changes in the initial configuration. However, such may not be the case for other partially ordered states and the evolution of the parameters $S$ and $\eta$ may be dependent on the initial condition. However, we can safely remark that the phase diagram shown for a particular initial state and at a particular time in Fig-9 is a typical one; in the sense that it is separated in three broad regimes corresponding to synchronization, incoherence and multicluster. Finding a complete description of the various fixed points of the system as well as their basins of attraction leading to various dynamical states would be an interesting future study worth pursuing.

 The study of how the system size $N$ affects the various dynamical states is also interesting. We already showed the results of simulations in detail for $N=100$ and $N=1000$, and the occurrence of the various dynamical states are expected at larger systems also. Many of the statistical properties of the system $1$ are dependent on initial conditions (even at fixed system size) due to its multistable nature, so it is hard to analyse the effects of change in system size on them. One of the more robust properties which we discuss about here is the time to reach complete synchronization. As is evident from the plot of the dynamics of the order parameter $R$ in Fig-12, synchronization occurs abruptly in systems of all sizes, but the time to reach that does not follow any obvious relation to the system size.

The phase plots in Fig-9 show some further interesting behaviour of the system. In both plots, a fine yet prominent band corresponding to states with exceptionally clustered phases are observed near $\epsilon=0.1$ for almost all, even small values of coupling radius which is surrounded by completely asynchronous states. The phase level population distribution plot in Fig-12 reveals this behaviour by showing a peak in population for some specific phase levels. Fig-12 also shows that this peakedness of the distribution occurs at a specific value of $\epsilon$ and is absent at other nearby values, where the almost uniform distribution implies incoherence.  As is evident from the phase plots (Fig-9), this behaviour is less prominent for larger values of the coupling radius.
 The mechanism for this exceptional clustering of phases at extremely small values of the coupling radius for a very specific coupling strength is yet unknown to us. However, using lattices of various sizes, and varied random initial conditions we have ensured that this behaviour is present in all cases, irrespective of lattice sizes and they occur at the same value of $\epsilon$ in all of them. More interestingly, for $l=50$, irrespective of lattice sizes, initial conditions and number of iteration steps, the oscillator phases always tend to take values mostly between two phase levels only, namely, the $10$th and the $20$th level (Fig-12) and this clustered state is static,i.e., after some transient dynamics, the automaton values do not change over time.

In the multicluster state of intermediate order, we have also investigated the distribution of sizes of synchronization clusters. As the defining property of a cluster in our context, we have demanded that the nearest-neighbor phase deviation $Z_{i}$ inside a cluster cannot have absolute value larger than $0$. With this definition, the clusters size distribution is seen to be a power law with exponent nearly equal to 1.5, as is shown in Fig-13. Furthermore, the power law exponent is independent of the values of the parameters $\epsilon$ and $r$ as is seen from Fig-13.

\noindent {\bf III. Summary and Future Extensions:}

In this article, we have developed a simple cellular automata model of 
nonlinearly coupled oscillators which reveals many essential features of
synchronization phenomena. In our simulation, we have observed that
depending on the values of coupling strength and radius of coupling we observed
various phases namely: asynchronous, where all the oscillators oscillate; synchronized, where all
oscillators oscillates coherently and other intermediate ordered states with spatially discontinuous profiles of the automata values. We have plotted the phase diagram in the plane described by strength of coupling and the radius of coupling using the
strength of incoherence. Furthermore, the partially ordered states 
were distinguished by discontinuity measure.

This model can be more generalized to include non-homogeneous coupling strengths and/or extended to other kinds
 (e.g. quadratic, cubic, broken linear etc.) of nonlinearity in the coupling term. The effect of time-delay in the coupling term can also be studied. The time needed by the system to organize into a chimera state starting from random initial conditions can be studied. Determination of the fate of the chimera state, i.e., whether or not the chimera state synchronizes globally after a sufficiently long time is also an interesting problem. Furthermore, one can also test the model in higher dimensions and other network geometries to explore its dynamical behaviour. The role of time dependent coupling, the hysteretic response (if any) and importantly, the existence of any competing time scale\cite{ba} may be the future objective of study in the cellular automata model of nonlinearly coupled oscillators.  

\newpage
\begin{figure}[h]
\begin{center}
\begin{tabular}{c}
        \resizebox{7cm}{!}{\includegraphics[angle=-90]{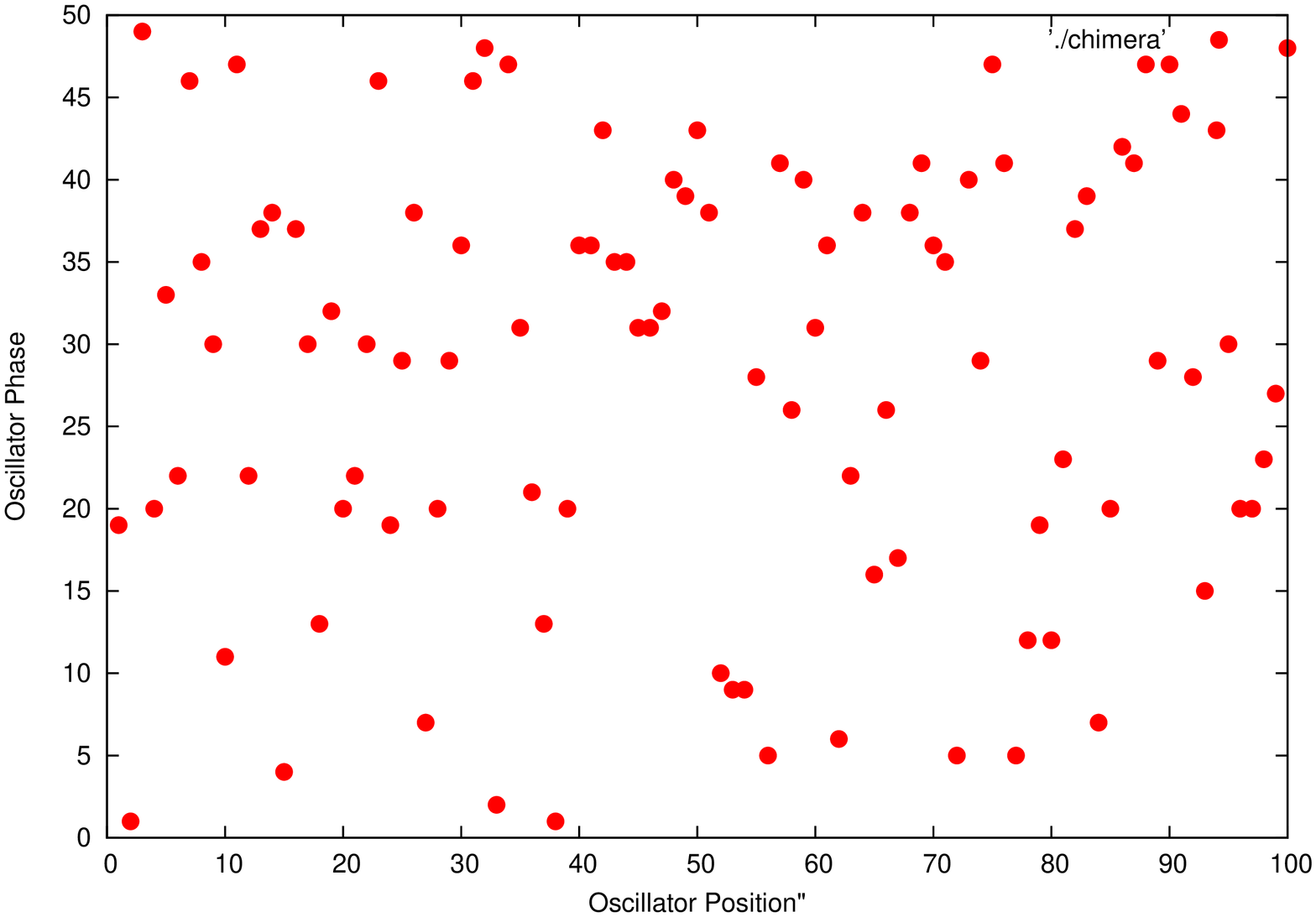}}
        \resizebox{7cm}{!}{\includegraphics[angle=-90]{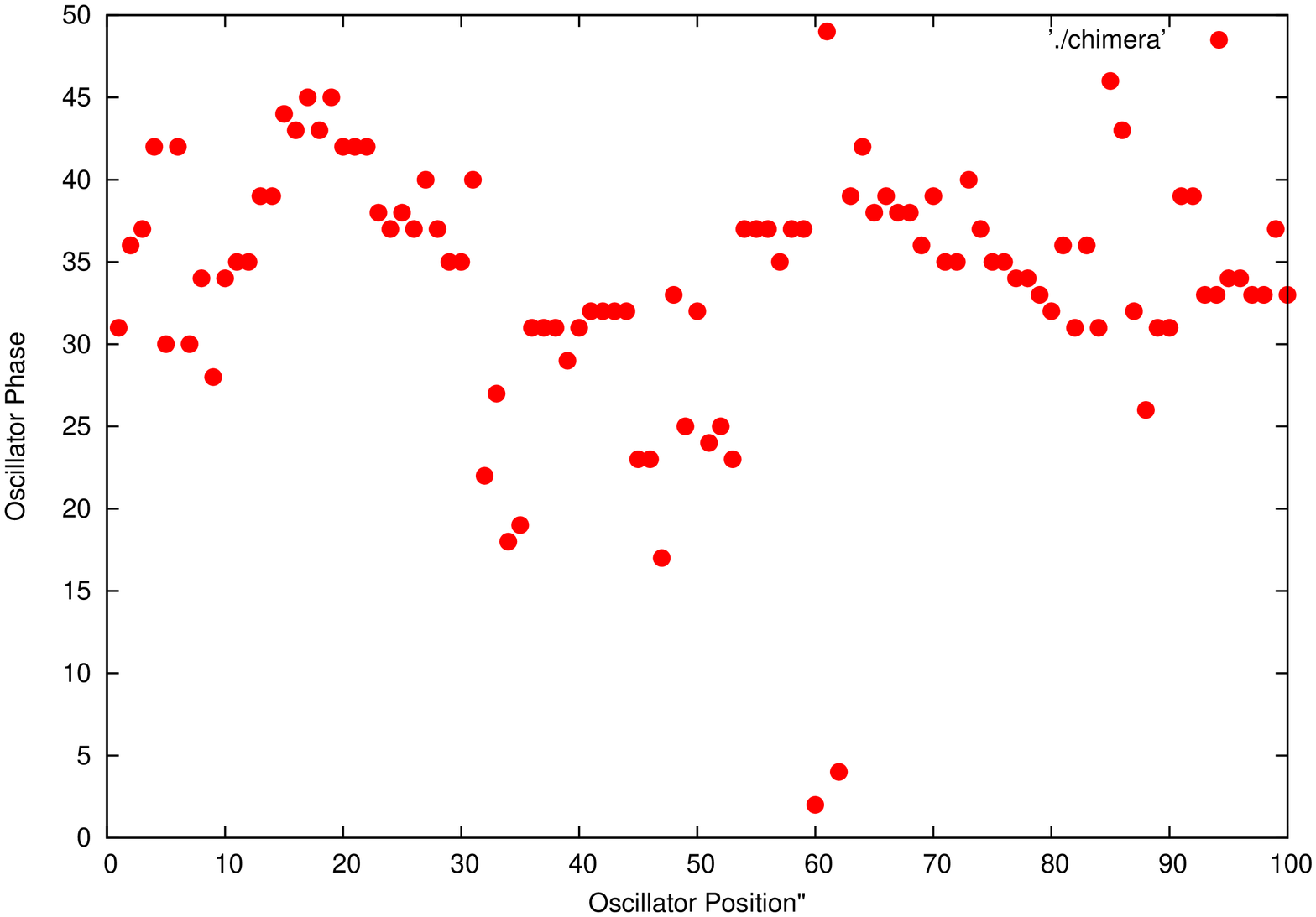}}
        \\
        \resizebox{7cm}{!}{\includegraphics[angle=-90]{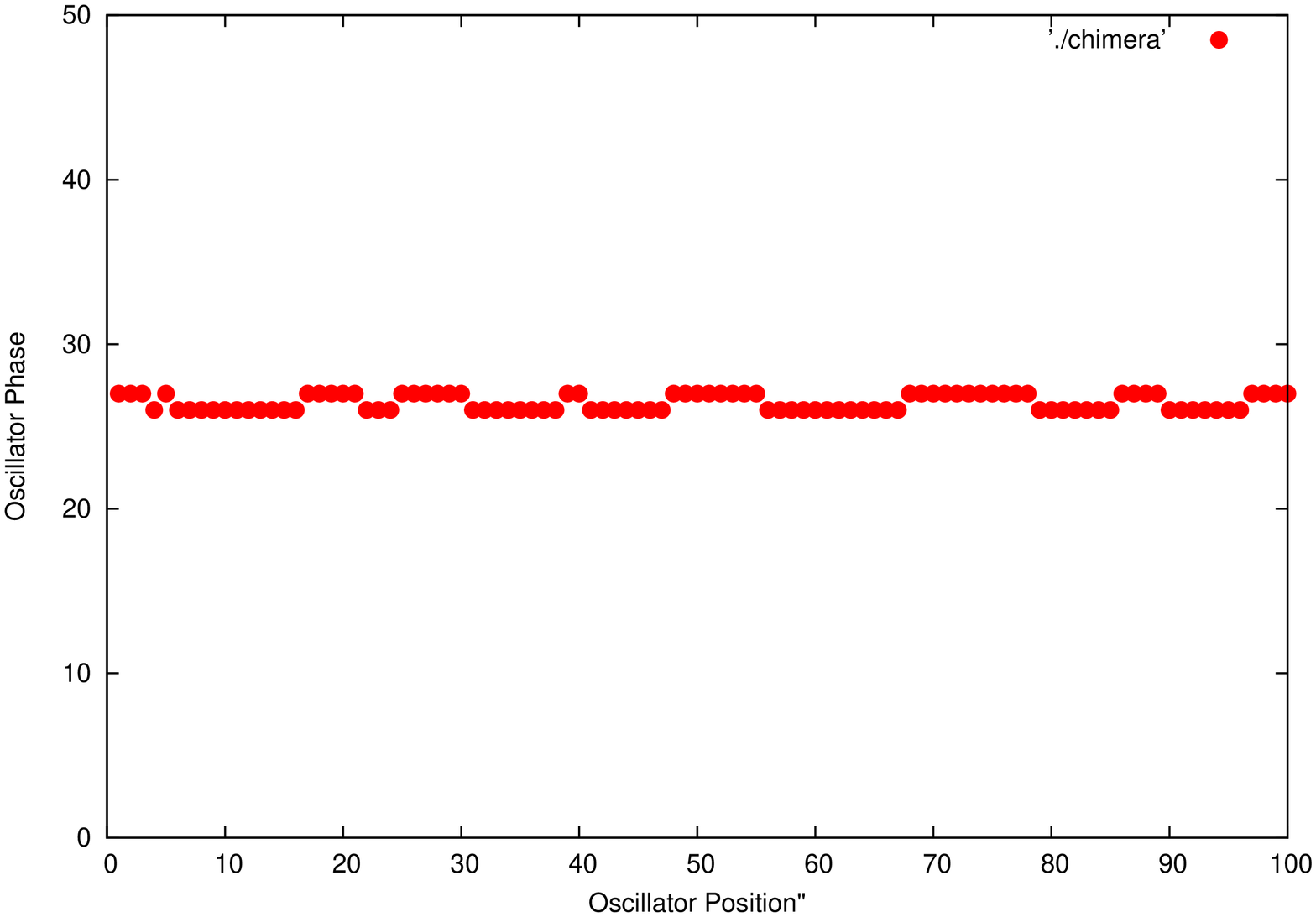}}
        
          \end{tabular}
 \caption{Snapshot of the Phases of $100$ Oscillators after $80000$ iterations revealing various Collective Dynamical States: (clockwise from top left) (a) Asynchronous State for $\epsilon=0.5, r=0.02$, (b) Multi-cluster State for $\epsilon=0.5, r=0.35$, and (c) Synchronized State for $\epsilon=0.8, r=0.45$.}
\end{center}
\end{figure}

\newpage

\begin{figure}[h]
\begin{center}
\begin{tabular}{c}
        \resizebox{7cm}{!}{\includegraphics[angle=-90]{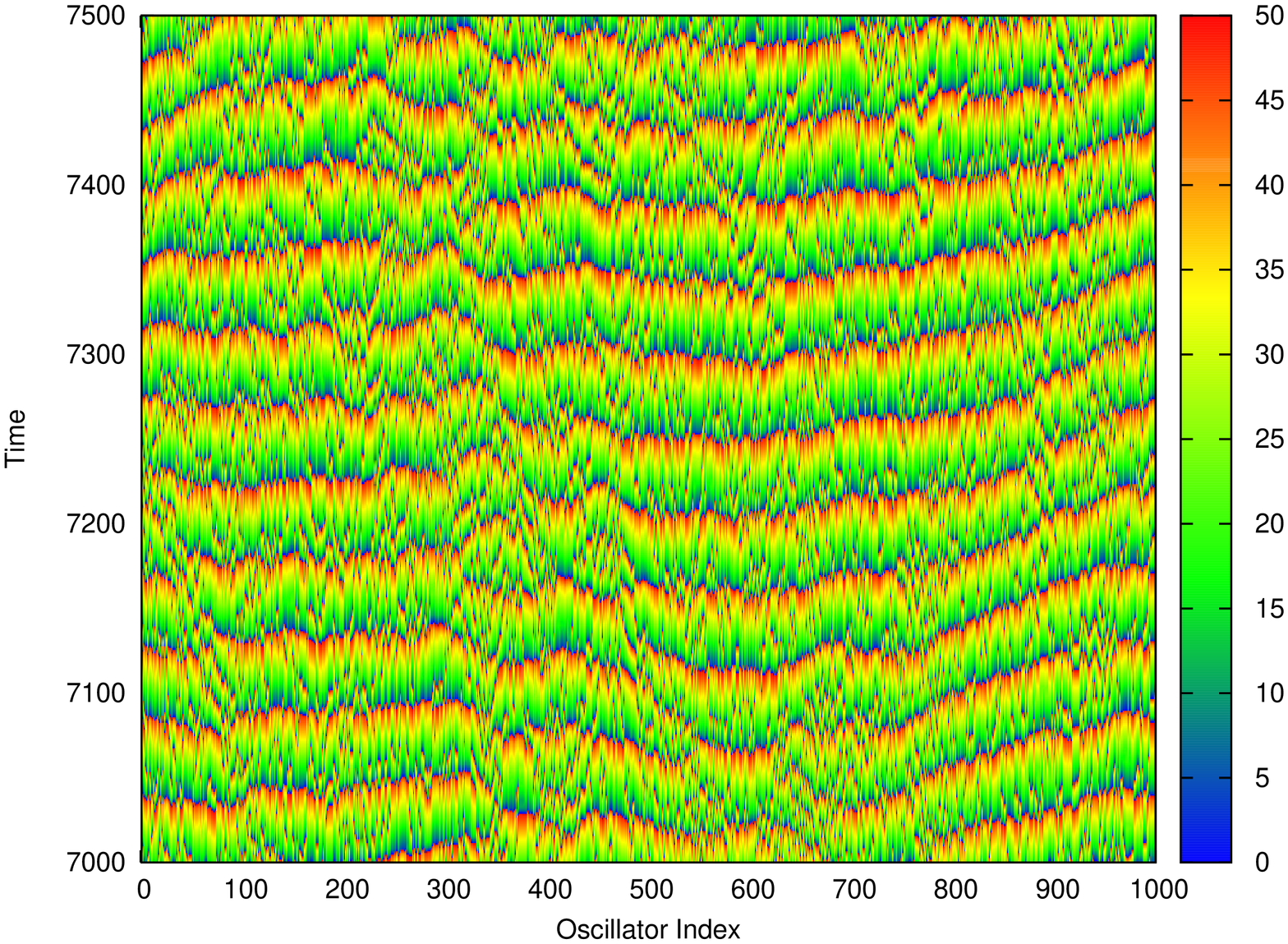}}
        \resizebox{7cm}{!}{\includegraphics[angle=-90]{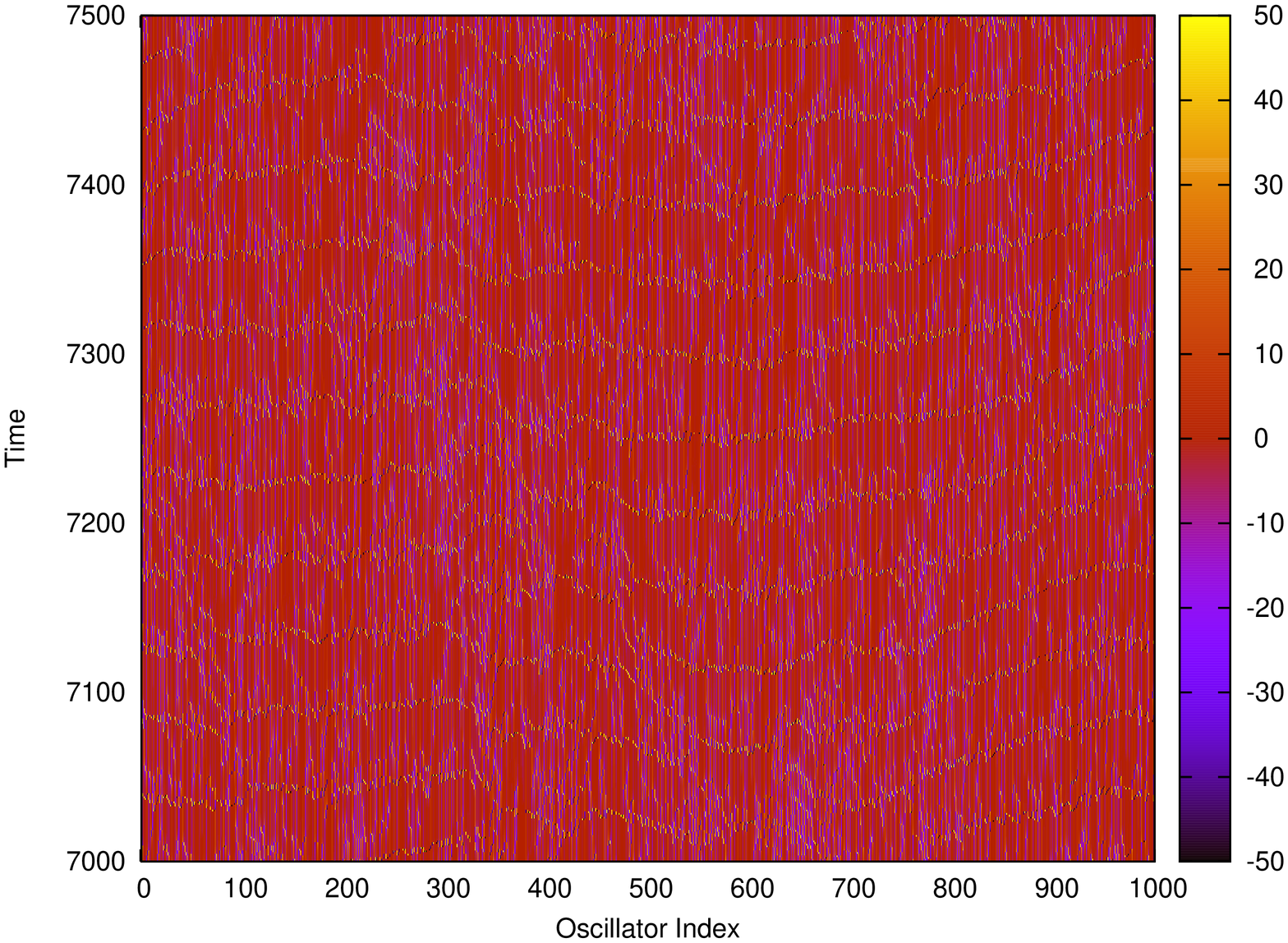}}
        \\
        \resizebox{7cm}{!}{\includegraphics[angle=-90]{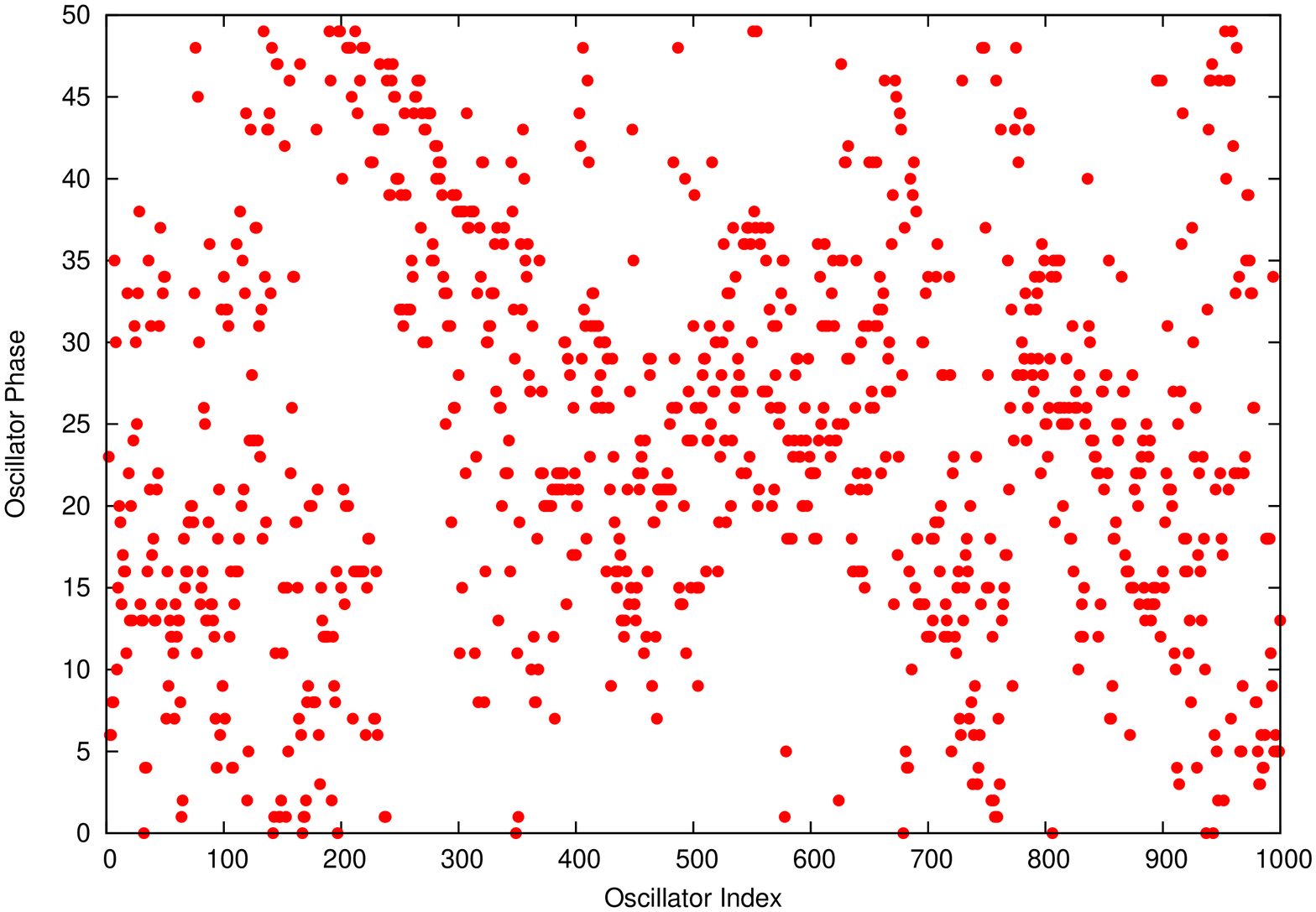}}
        
          \end{tabular}
 \caption{Dynamics of The Asynchronous State with $L=1000$,$r=0.02$,$\epsilon=0.2$: The Space-Time Dynamics of the Phases (top left), Nearest Neighbour Phase Differences (top right) and a Particular Snapshot at Iteration Step $7001$ (bottom).}
\end{center}
\end{figure}
\newpage
\begin{figure}[h]
\begin{center}
\begin{tabular}{c}
        \resizebox{7cm}{!}{\includegraphics[angle=-90]{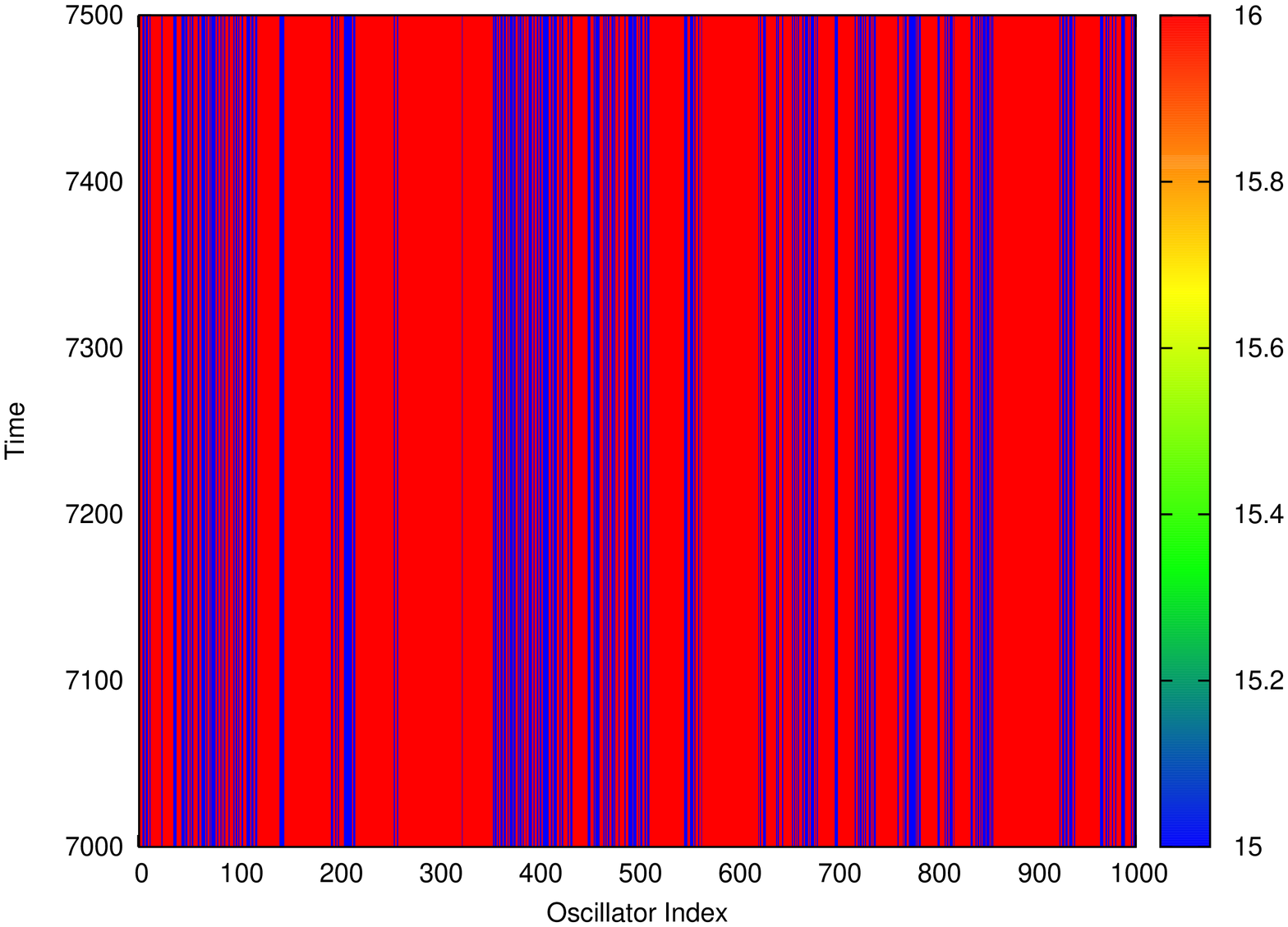}}
        \resizebox{7cm}{!}{\includegraphics[angle=-90]{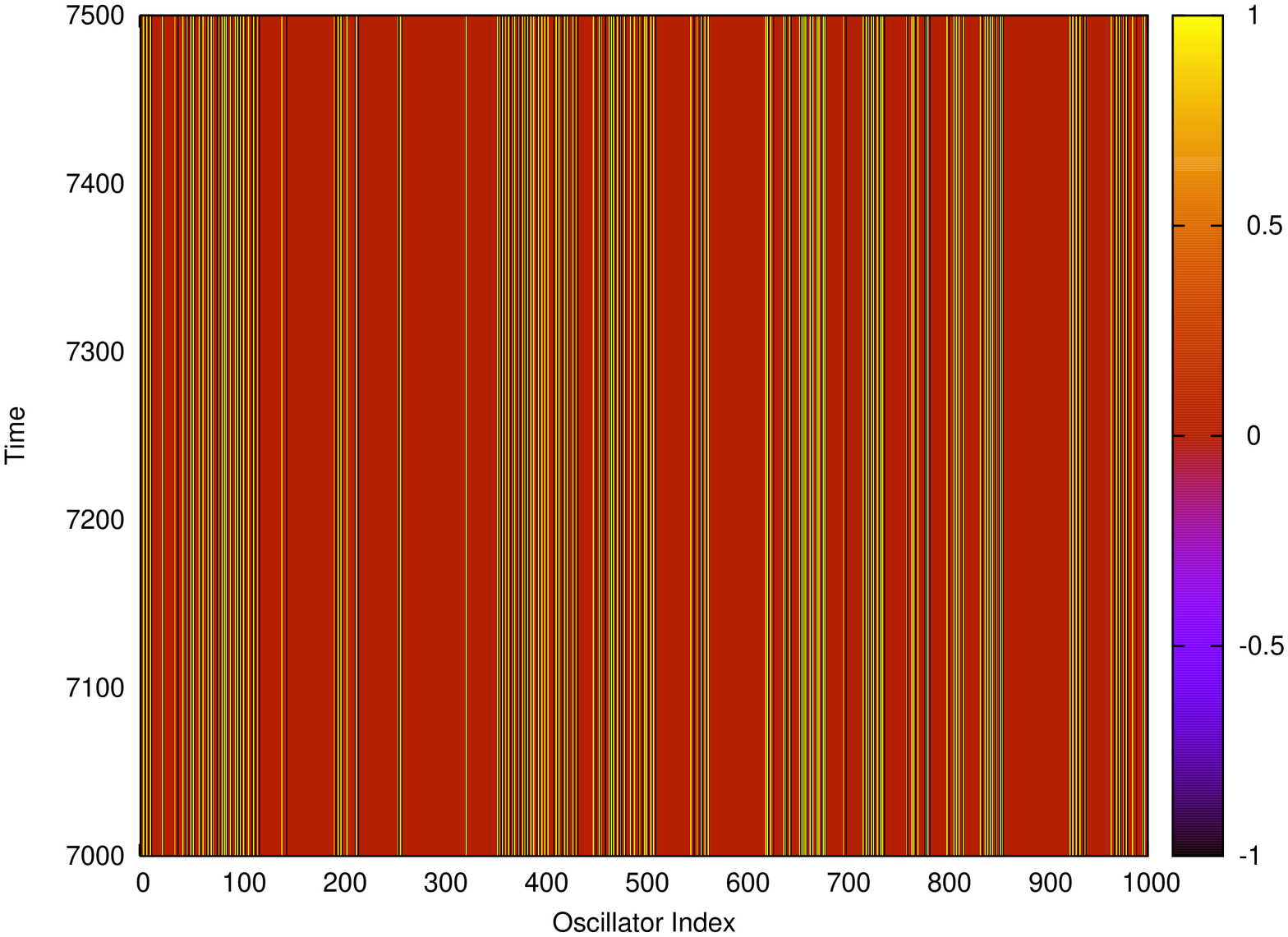}}
        \\
        \resizebox{7cm}{!}{\includegraphics[angle=-90]{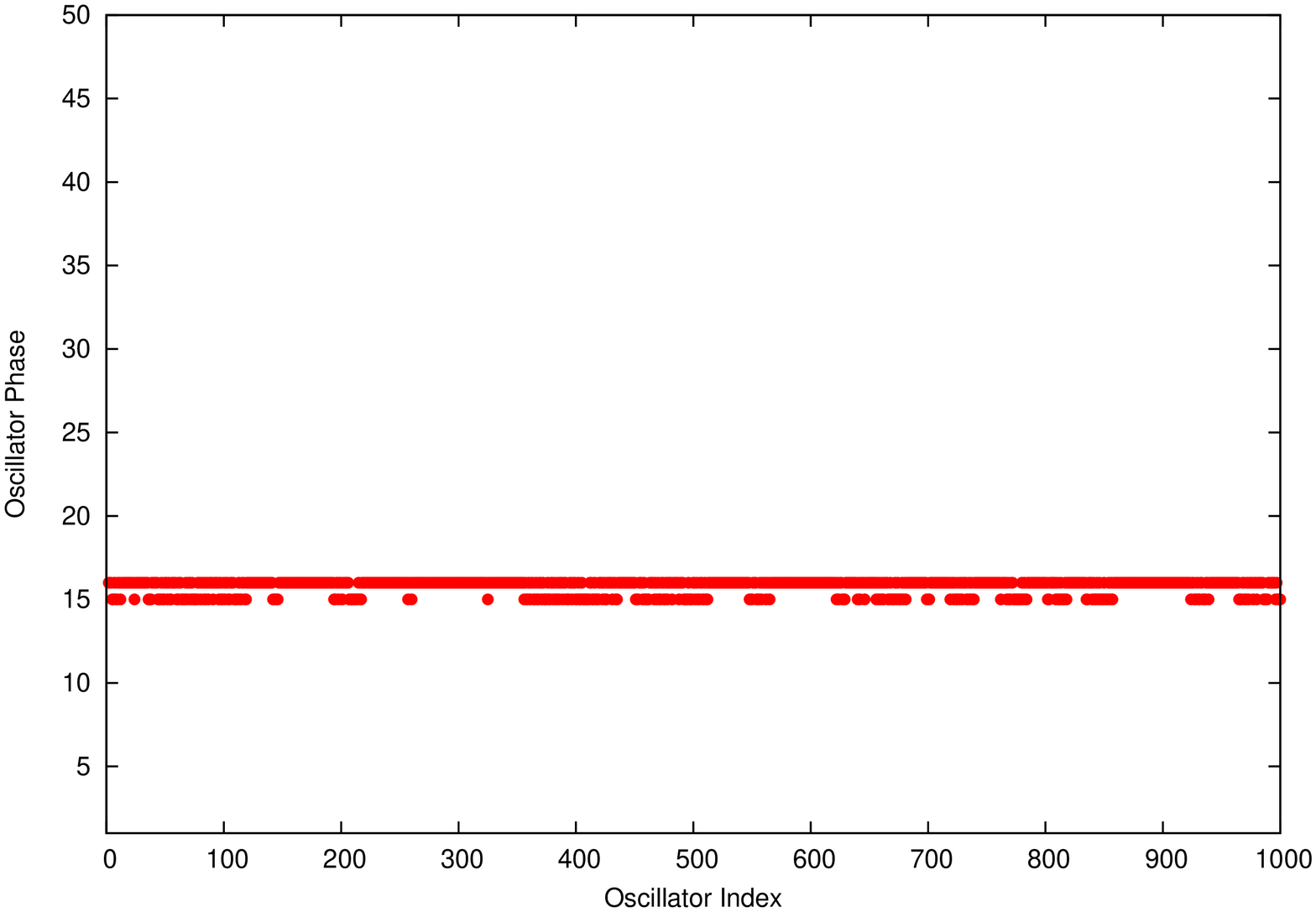}}
        
          \end{tabular}
 \caption{Dynamics of The Synchronous State with $L=1000$,$r=0.49$,$\epsilon=1.0$: The Space-Time Dynamics of the Phases (top left), Nearest Neighbour Phase Differences (top right) and a Particular Snapshot at Iteration Step $7001$ (bottom).}
\end{center}
\end{figure}

\newpage
\begin{figure}[h]
\begin{center}
\begin{tabular}{c}
        \resizebox{7cm}{!}{\includegraphics[angle=-90]{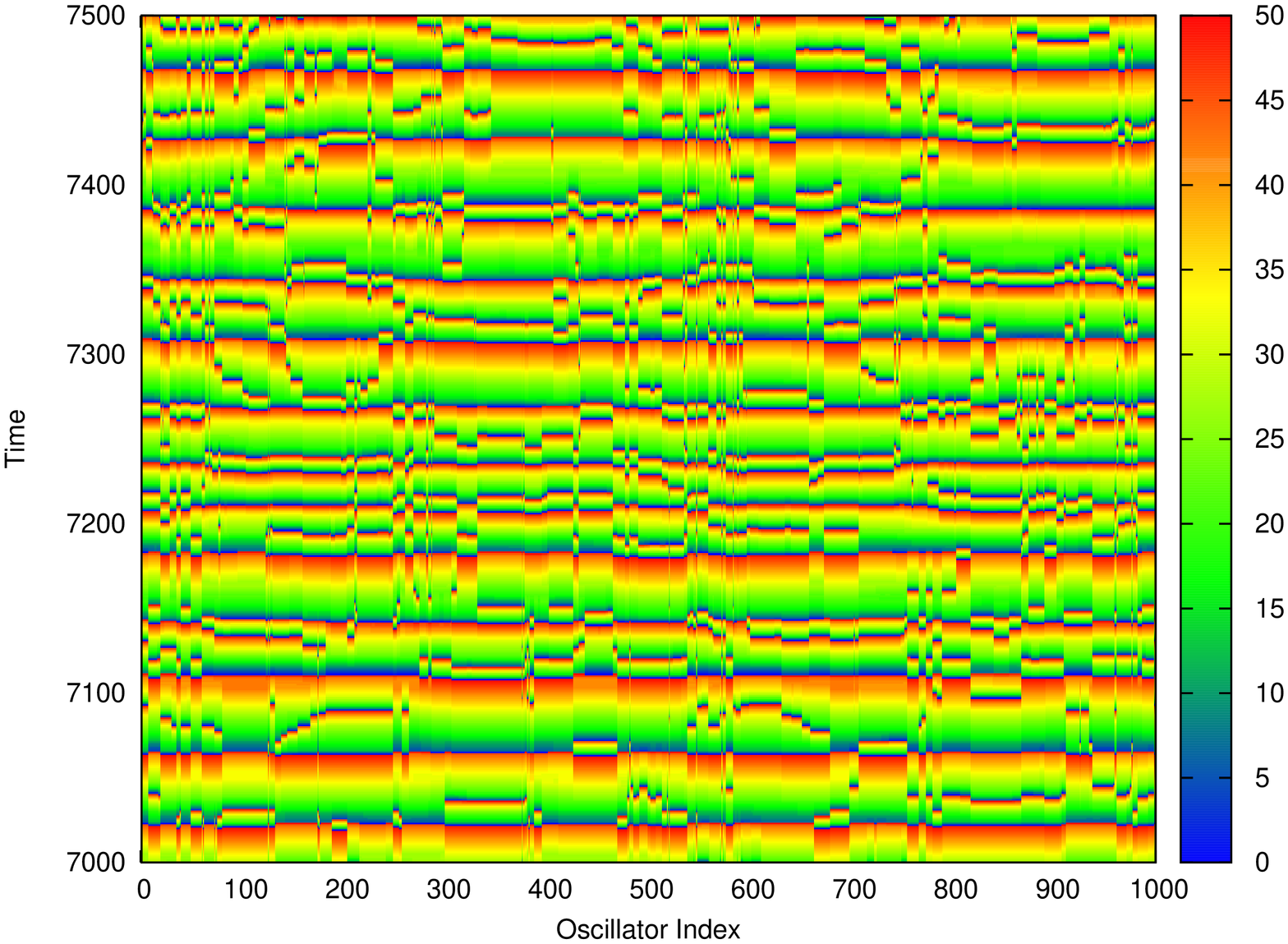}}
        \resizebox{7cm}{!}{\includegraphics[angle=-90]{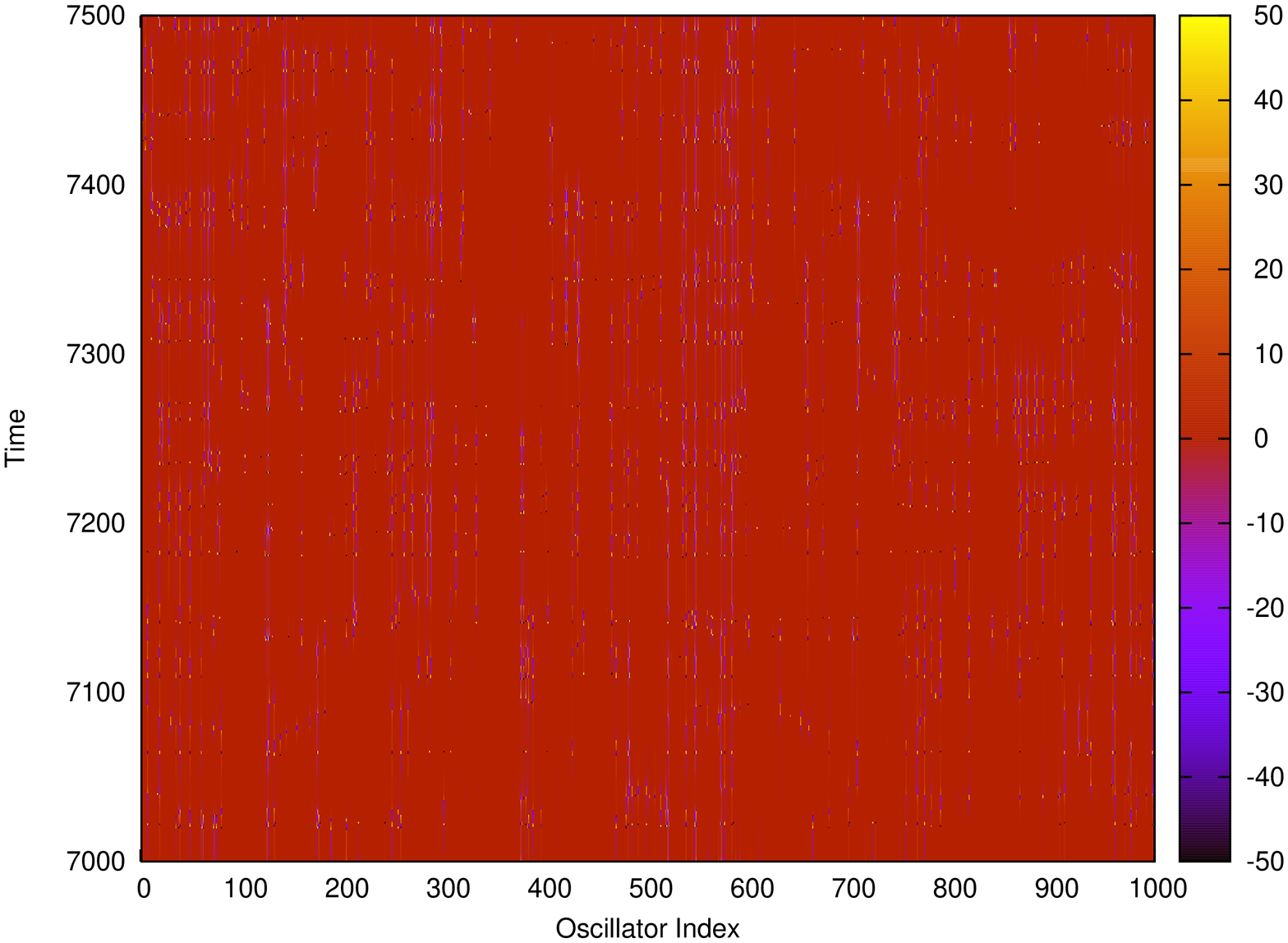}}
        \\
        \resizebox{7cm}{!}{\includegraphics[angle=-90]{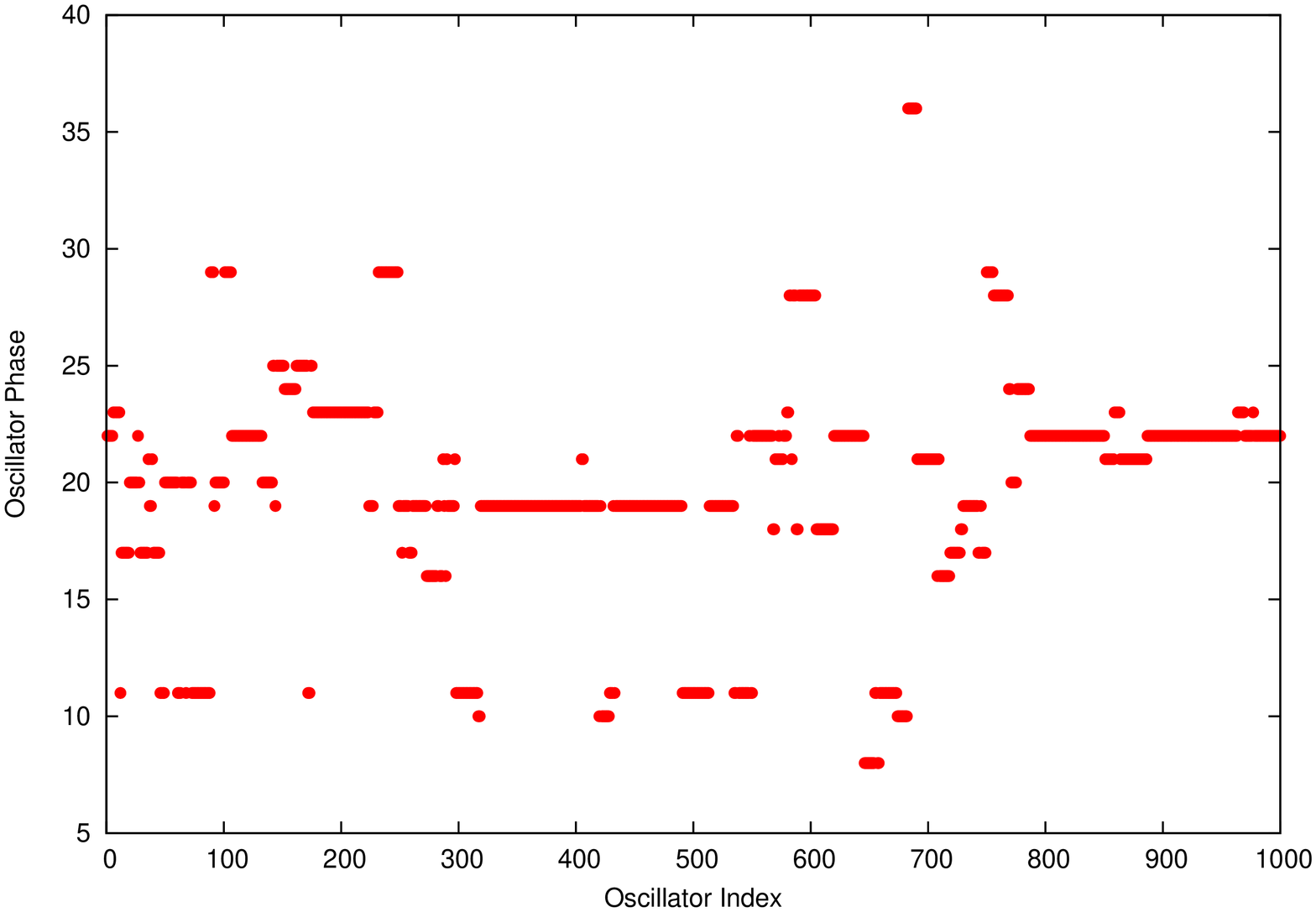}}
        
          \end{tabular}
 \caption{Dynamics of The partially ordered State with $L=1000$,$r=0.454$,$\epsilon=0.3$: The Space-Time Dynamics of the Phases (top left), Nearest Neighbour Phase Differences (top right) and a Particular Snapshot at Iteration Step $7398$ (bottom).}
\end{center}
\end{figure}
\newpage
\begin{figure}[h]
\begin{center}
\begin{tabular}{c}
        \resizebox{8cm}{!}{\includegraphics[angle=-90]{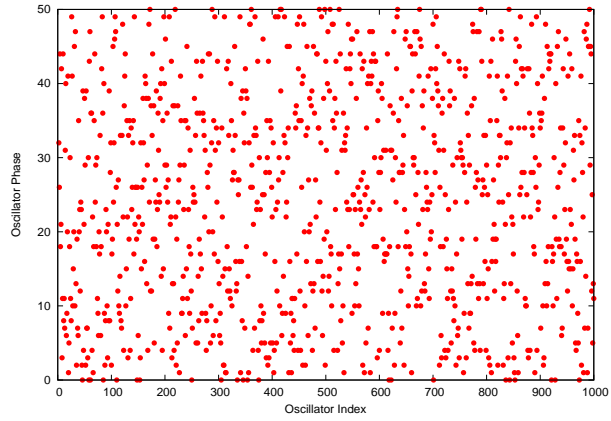}}
                
          \end{tabular}
 \caption{The Initial Random Phase Distribution from which all the states shown in the last $3$ plots are evolving.}
\end{center}
\end{figure}
\newpage
\begin{figure}[h]
\begin{center}
\begin{tabular}{c}
        \resizebox{12cm}{!}{\includegraphics[angle=-90]{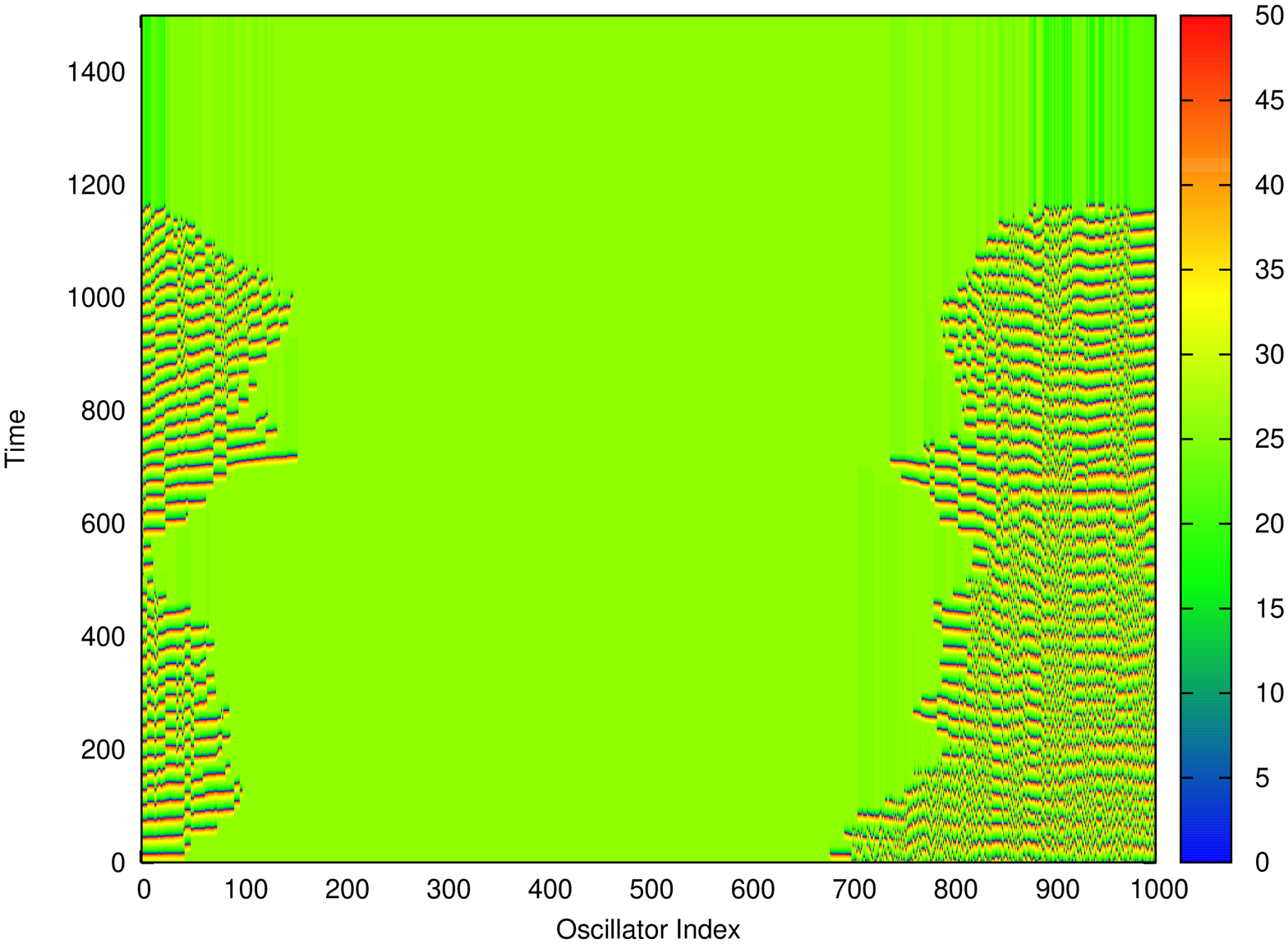}}\\
	\resizebox{12cm}{!}{\includegraphics[angle=-90]{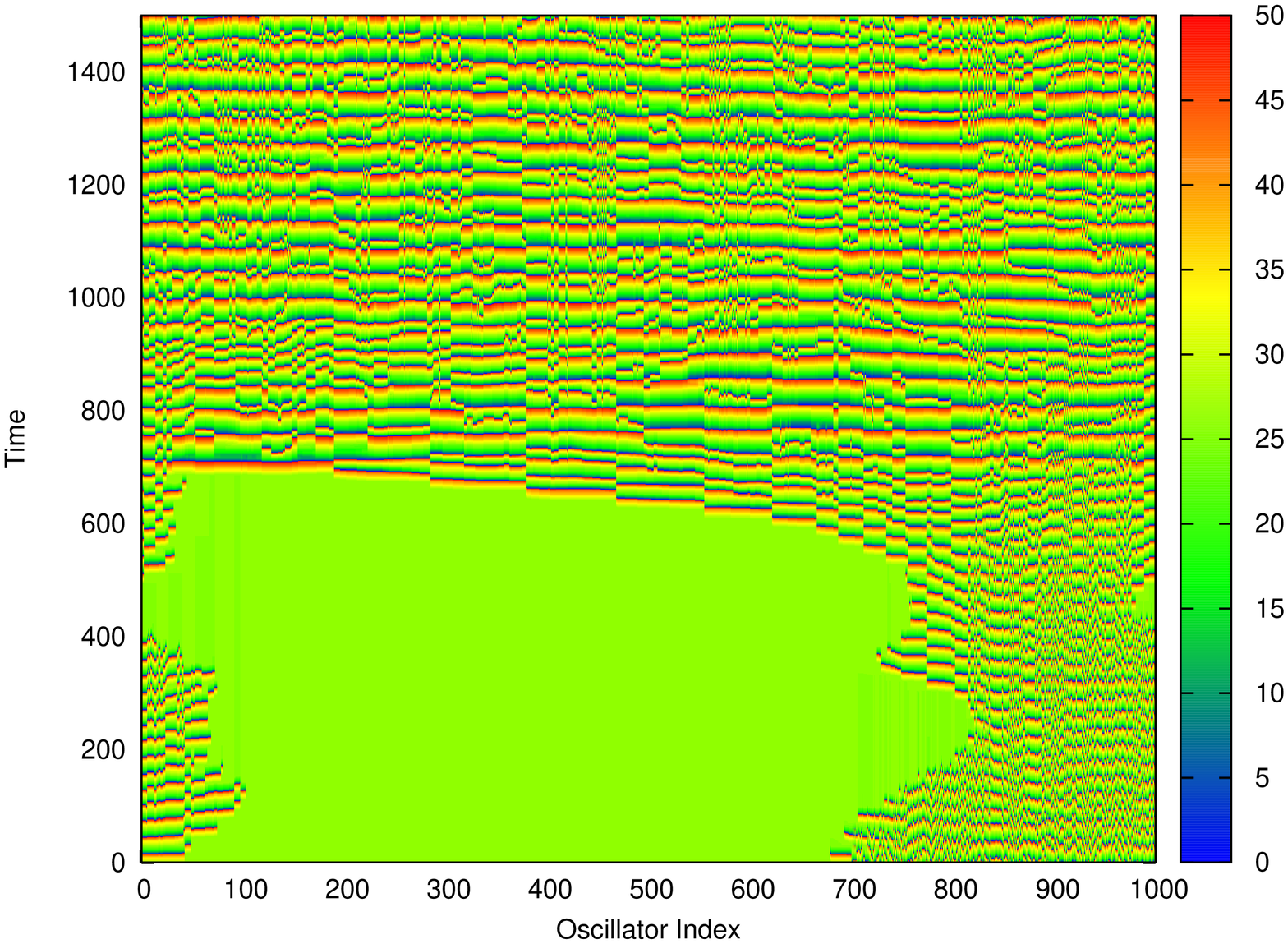}}
        
          \end{tabular}
 \caption{Space-Time Plot for Chimera States for $N=1000$, $N_c=149$,$n=700$, $l=50$ and $c=26$ with $\epsilon =0.192405$(top) and $\epsilon=0.1924236$ (bottom), Note the change in dynamical behaviour with the Slight change in $ \epsilon $.}
\end{center}
\end{figure}

\newpage
\begin{figure}[h]
\begin{center}
\begin{tabular}{c}
        \resizebox{10cm}{!}{\includegraphics[angle=0]{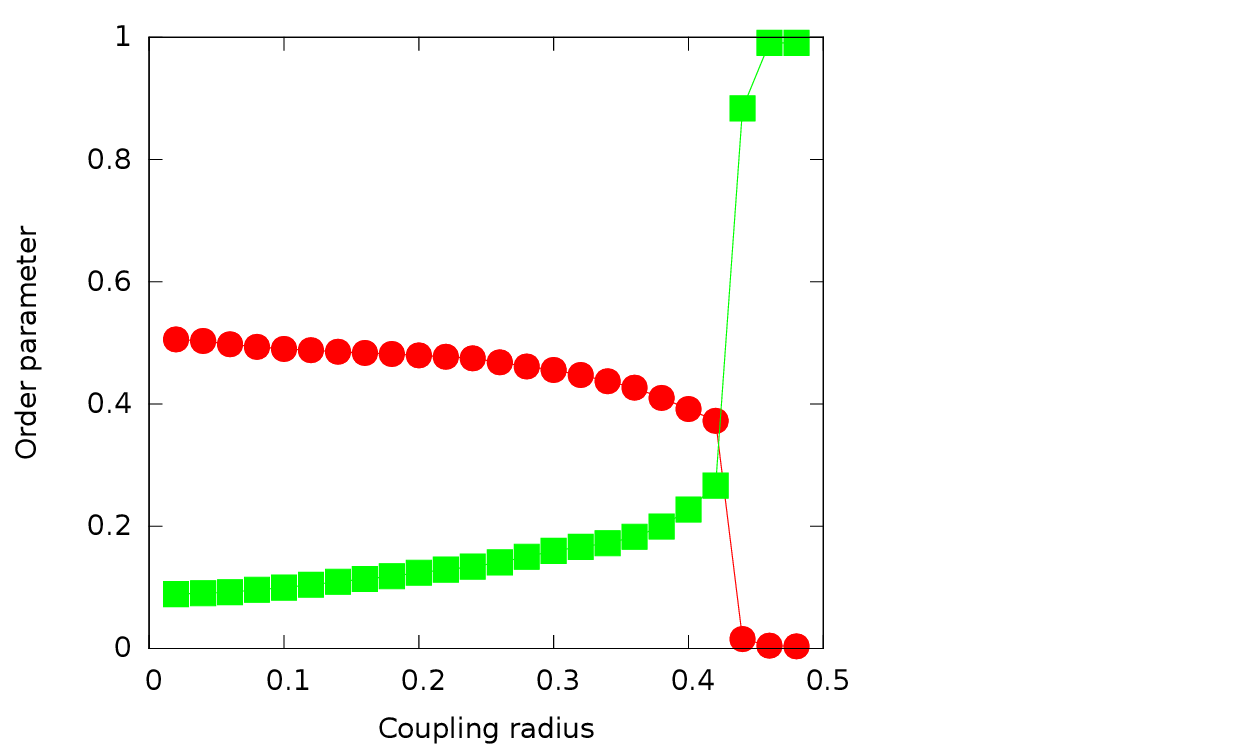}}
                  \end{tabular}
 \caption{Sharp Change of the Order Parameters $R$ (box) and Normalized Standard Deviation $R'$ (circle) with the variation of Coupling Radius showing Synchronizing Transition. Here $N=100$,$\epsilon=1.0$ and the order parameters are averaged between times 40000 to 80000.}
\end{center}
\end{figure}
\newpage
\begin{figure}[h]
\begin{center}
\begin{tabular}{c}
        \resizebox{7cm}{!}{\includegraphics[angle=-90]{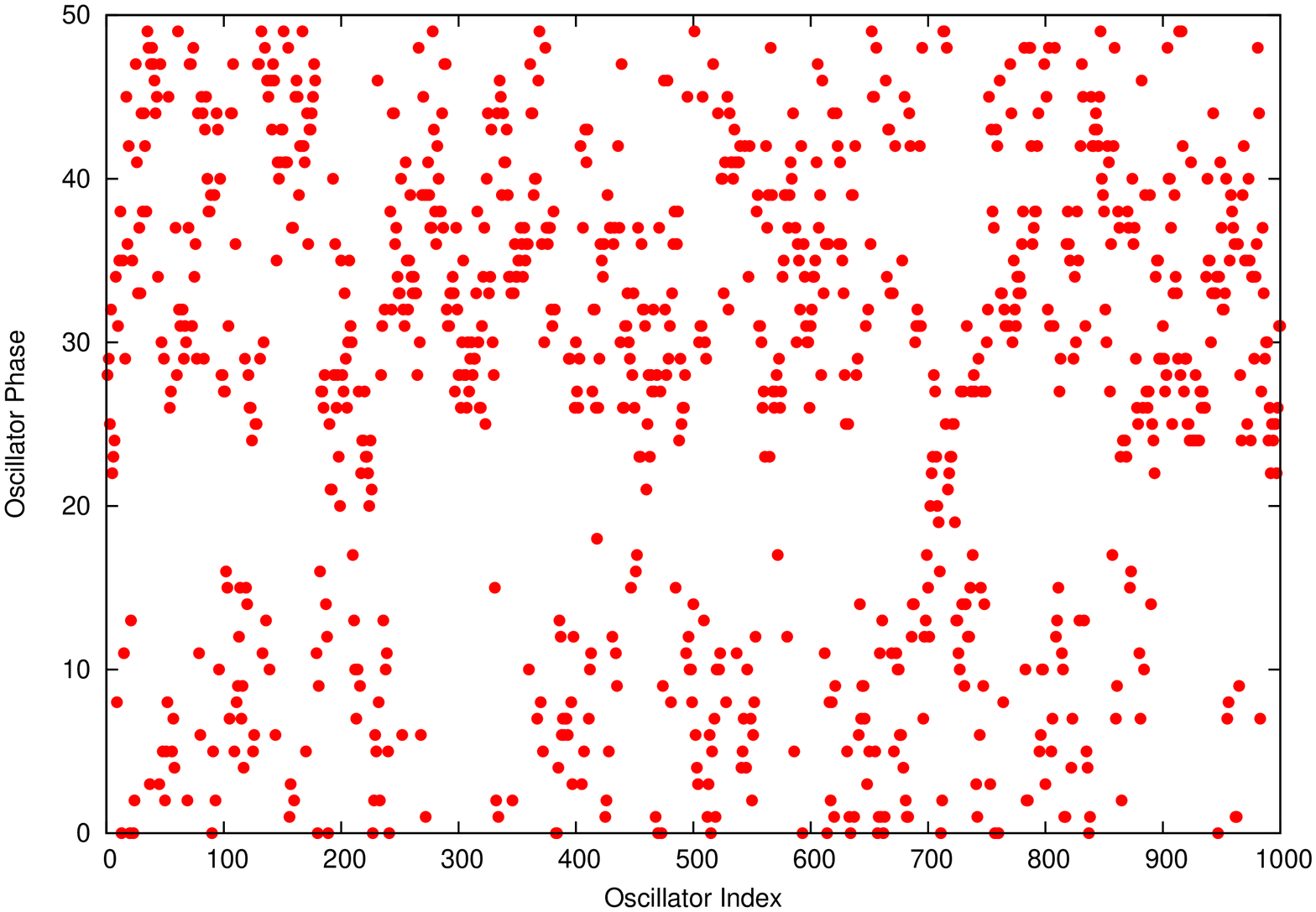}}
        \resizebox{7cm}{!}{\includegraphics[angle=-90]{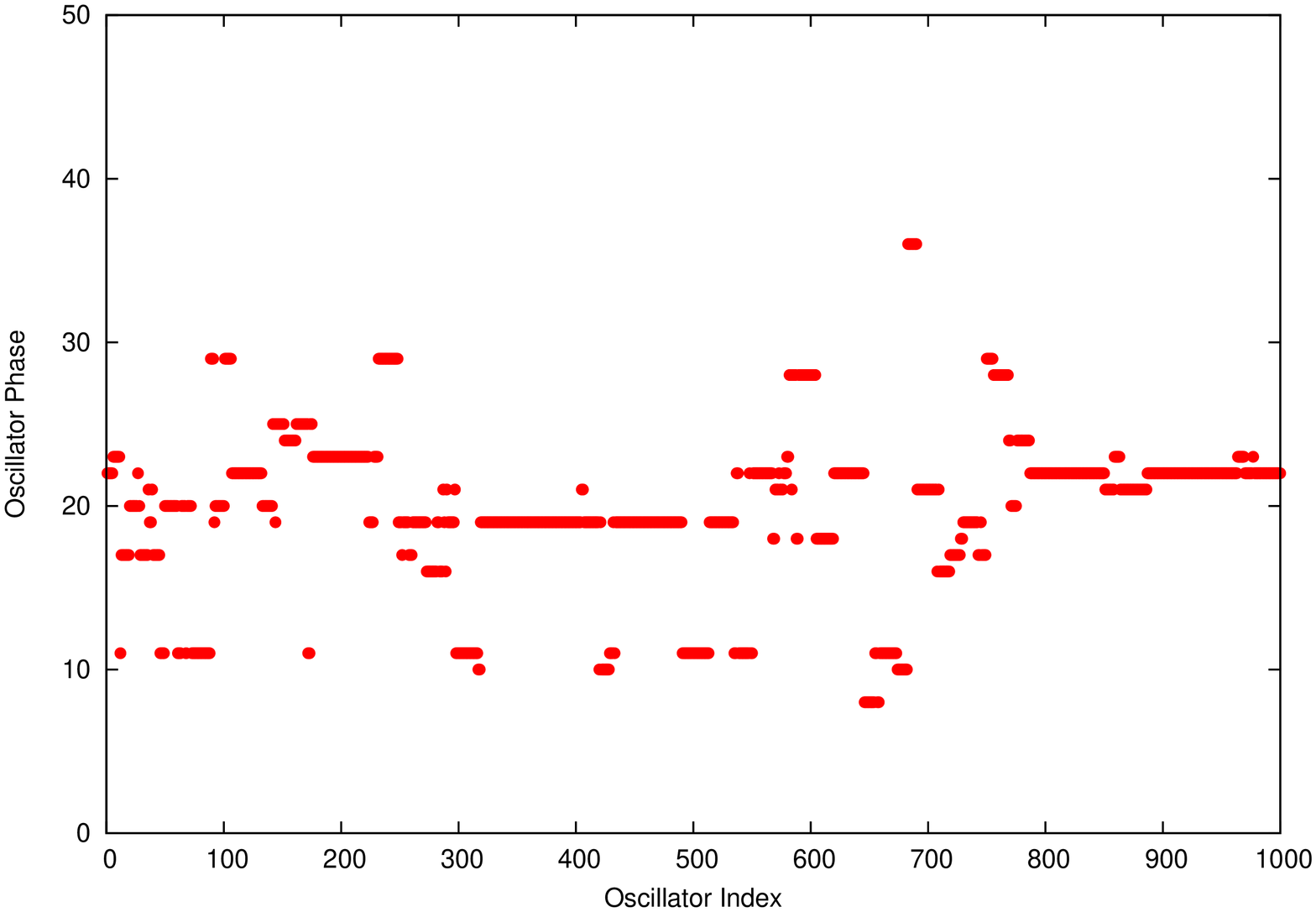}}
        \\
        \resizebox{7cm}{!}{\includegraphics[angle=-90]{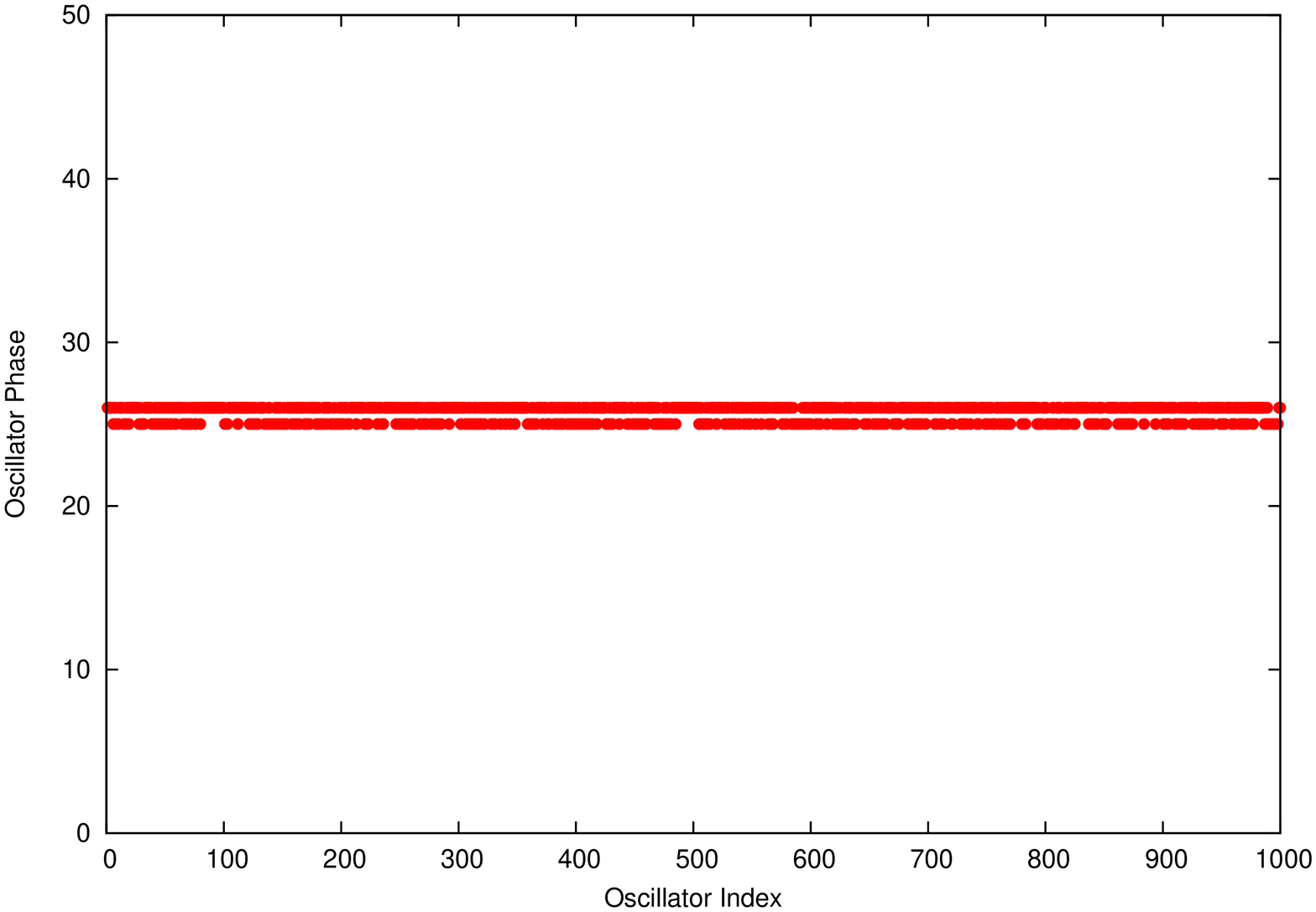}}
        \resizebox{7cm}{!}{\includegraphics[angle=-90]{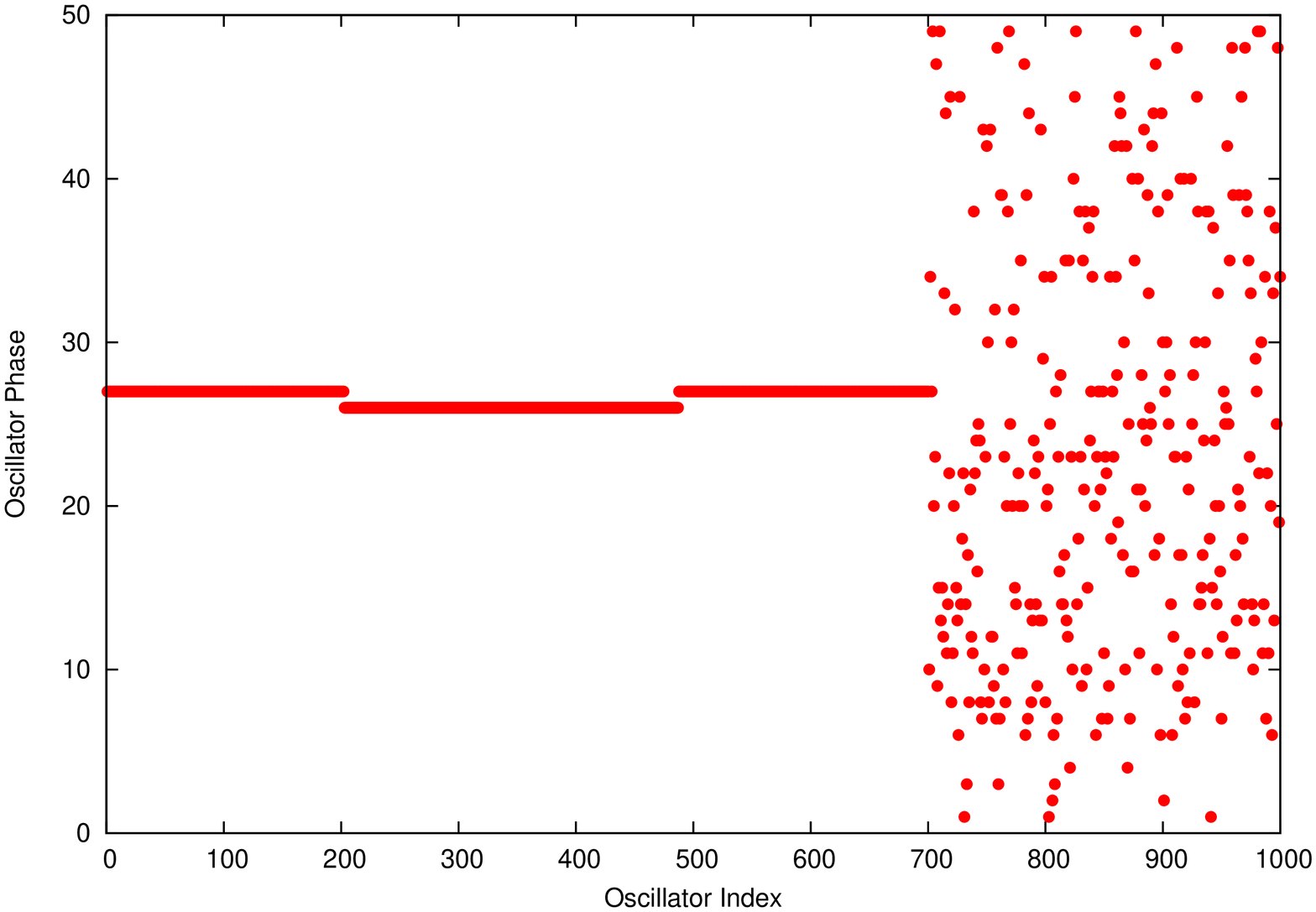}}
        
          \end{tabular}
 \caption{Characterization of Dynamical States using Strength of Incoherence $S$ and Discontinuity Measure $\eta$: (clockwise from top Left) Asynchronous ($S=1,\eta=0$), Multicluster ($S=0.5,\eta=2$), Chimera ($S=0.3,\eta=1$) and Globally Synchronized ($S=0,\eta=0$) States. Here we have taken $L=1000$, $N_g=10$ and $\delta=2.0$. Note that in the Chimera and Multichimera States, the value of $\eta$ is roughly equal to the number of the synchronized clusters present in the system.}
\end{center}
\end{figure}
\newpage
\begin{figure}[h]
\begin{center}
\begin{tabular}{c}
        \resizebox{7cm}{!}{\includegraphics[angle=0]{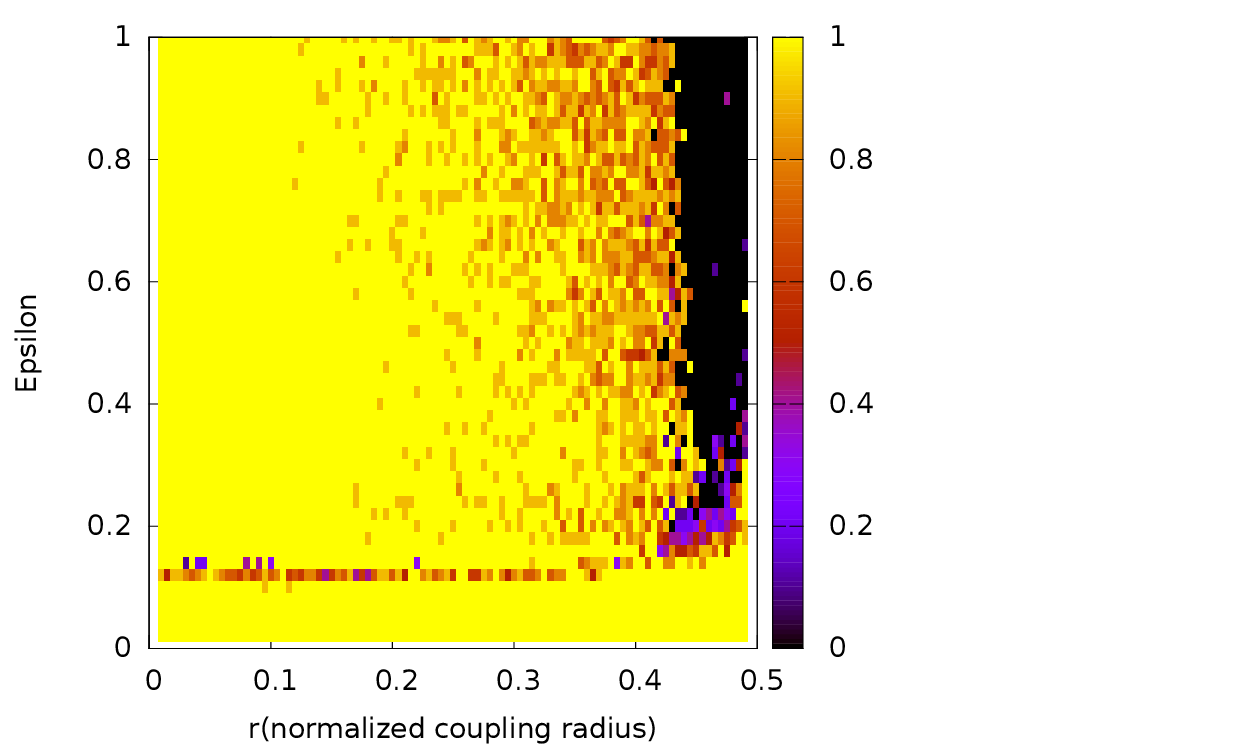}}
        \resizebox{7cm}{!}{\includegraphics[angle=0]{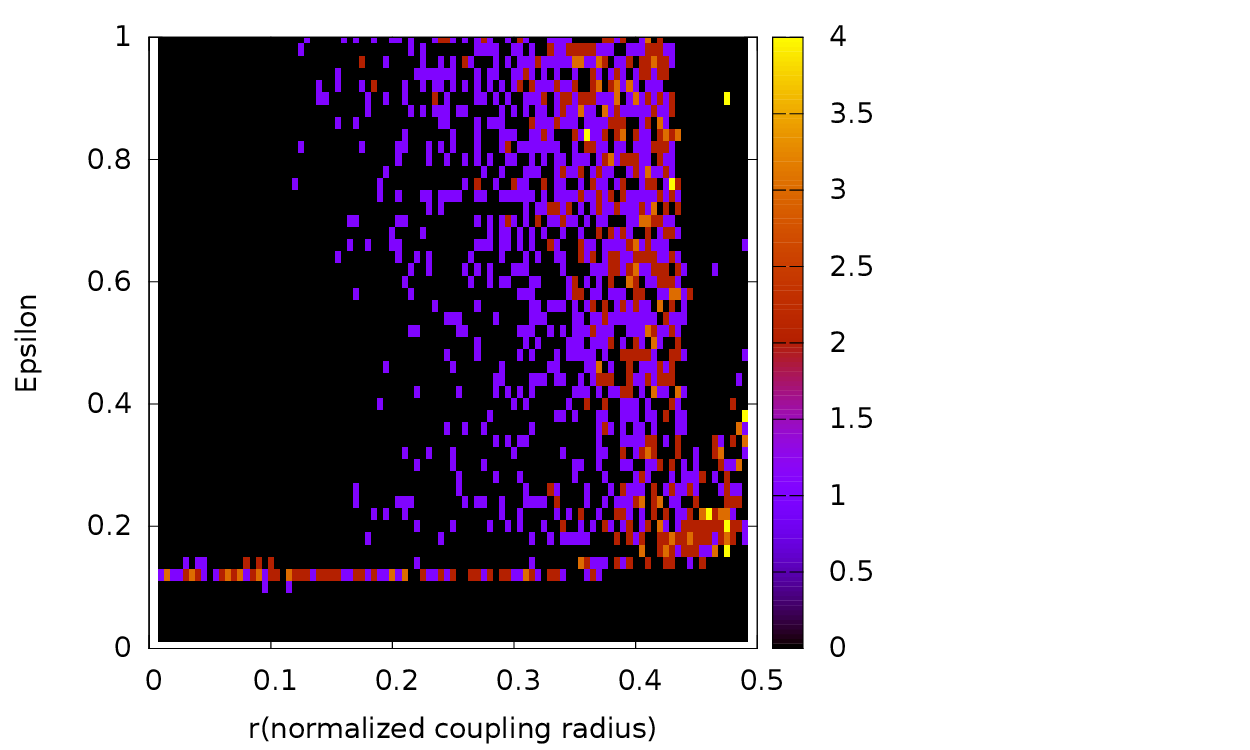}}
         \\
        \resizebox{7cm}{!}{\includegraphics[angle=-90]{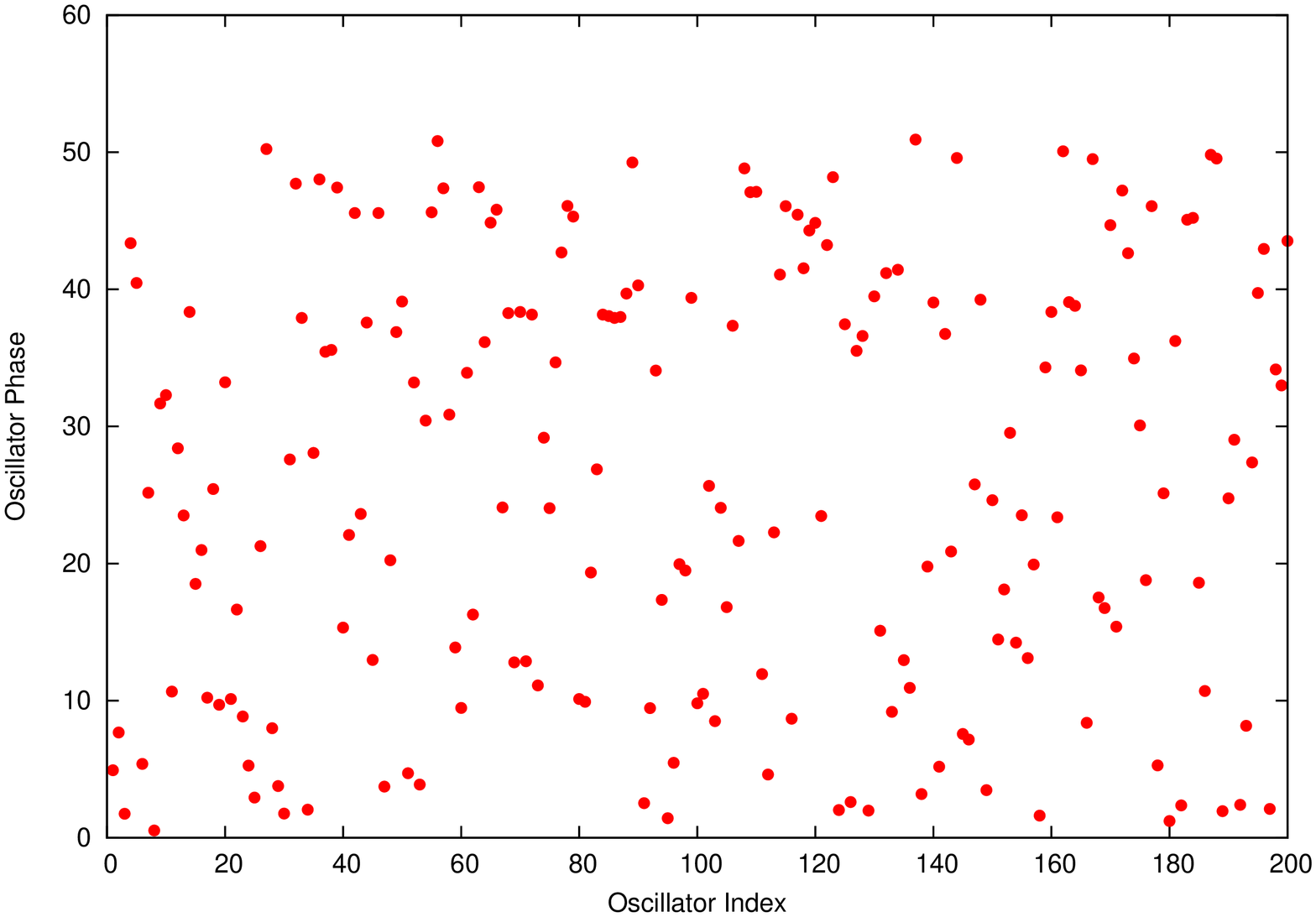}}
        
          \end{tabular}
 \caption{Phase Diagrams of $200$ Oscillators distinguishing between various Collective Dynamical States: (a) Strength of Incoherence (top left) and (b) Discontinuity Measure (top right). To generate the plots, we have taken $l=50$, $\delta=2.0$ and $N_{g}=10$ and the measures are done after $40000$ iterations starting from the random configuration (bottom).}
\end{center}
\end{figure}
\newpage
\begin{figure}[h]
\begin{center}
\begin{tabular}{c}
        \resizebox{10cm}{!}{\includegraphics[angle=-90]{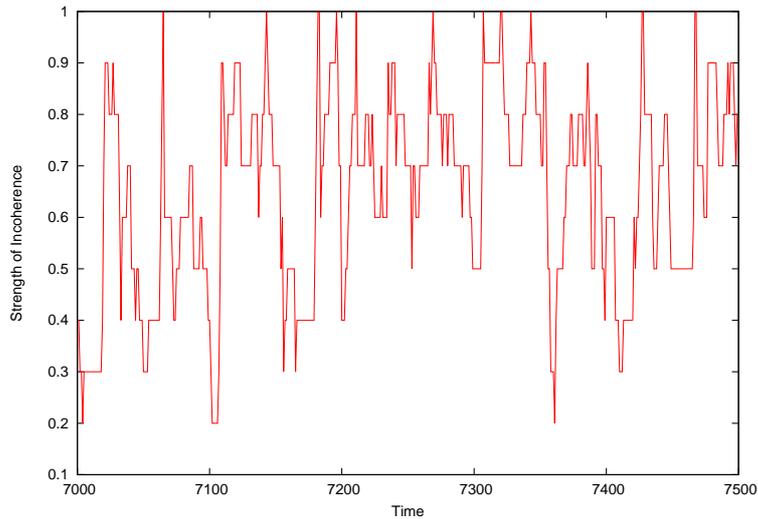}}
                  \end{tabular}
 \caption{The Time Variation of Strength of Incoherence for the partially ordered state shown in Fig-, with $\delta=2.0$ and $N_{g}=10$.}
\end{center}
\end{figure}

\newpage
\begin{figure}[h]
\begin{center}
\begin{tabular}{c}
        
        \resizebox{10cm}{!}{\includegraphics[angle=-90]{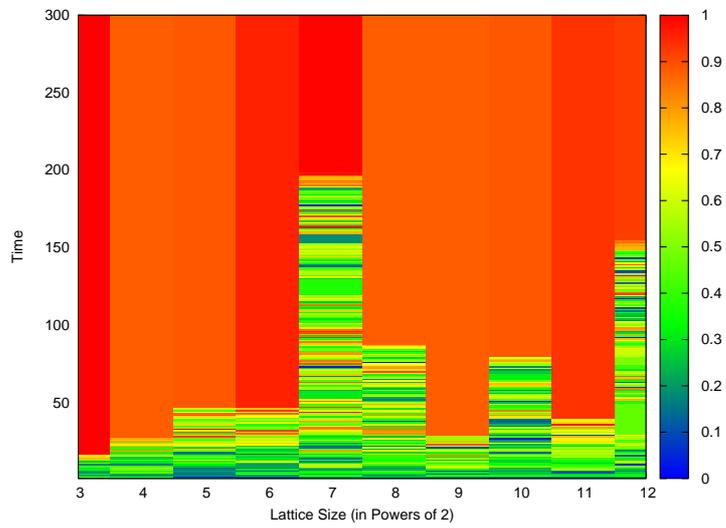}}
        
          \end{tabular}
 \caption{The Dynamics of the Order Parameter $R$ in Systems of Various Sizes, with $l=50$, $\epsilon=1.0$ and $r=0.49$.}
\end{center}
\end{figure}
\newpage
\begin{figure}[h]
\begin{center}
\begin{tabular}{c}
        \resizebox{12cm}{!}{\includegraphics[angle=-90]{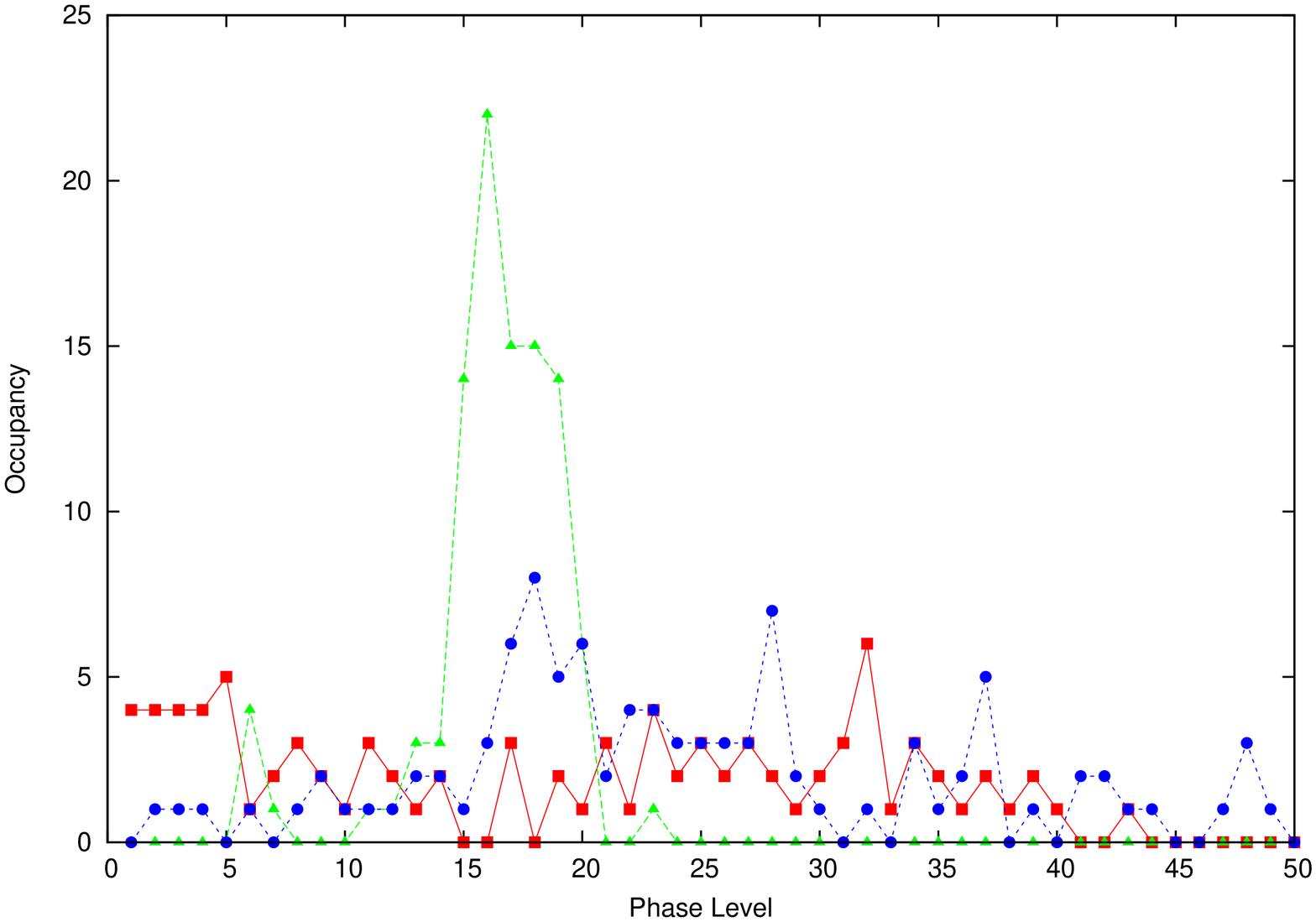}}
         \\
        \resizebox{12cm}{!}{\includegraphics[angle=-90]{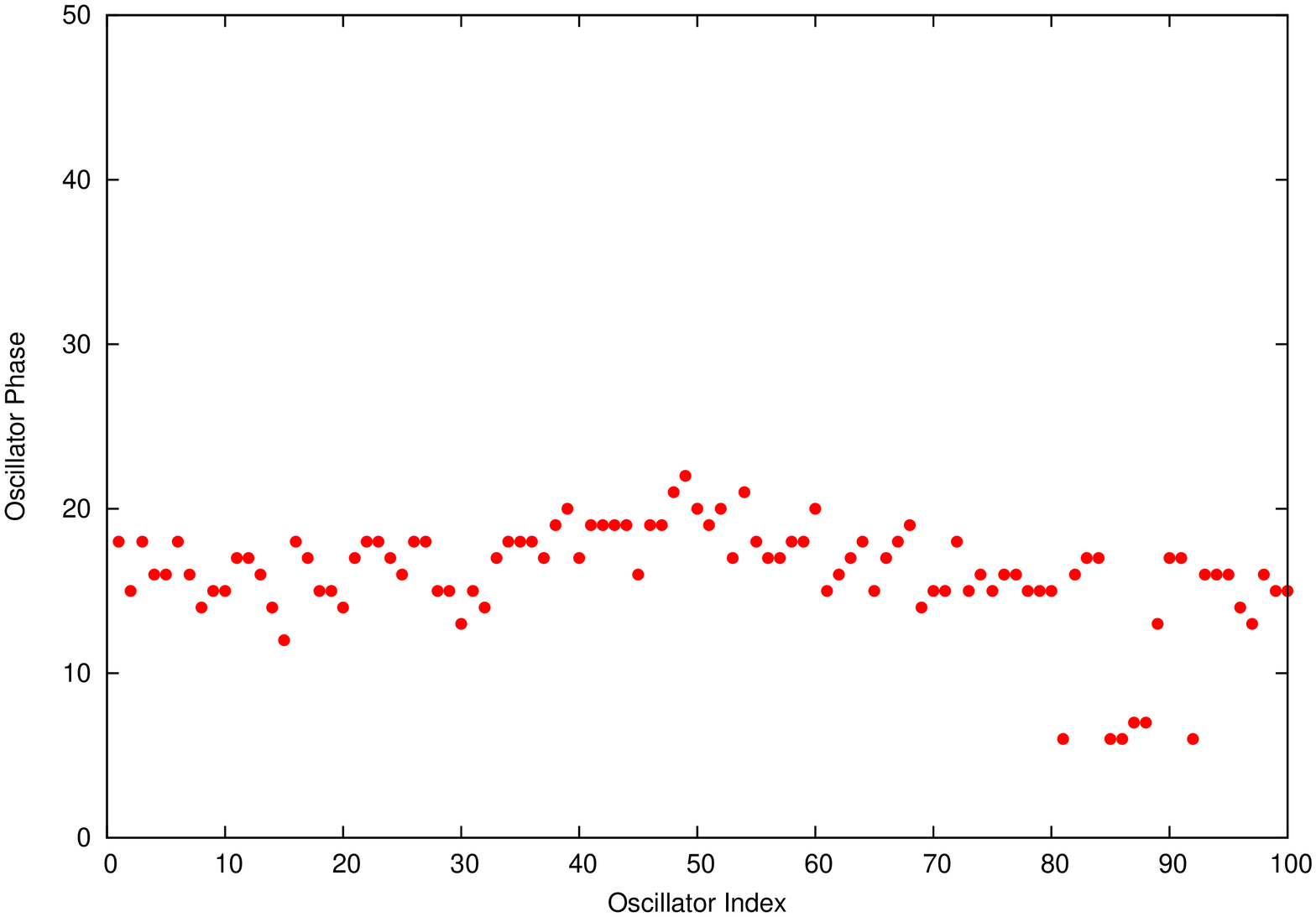}}
        
          \end{tabular}
 \caption{(Top) Phase Level population Distributions for coupling strengths $\epsilon=0.05$ (boxes), $\epsilon=0.12$ (triangles) and $\epsilon=0.2$ (circles) showing the exceptional cluster formation at a specific value of the coupling strength, represented by a sharp peak in the distribution. The snapshot of the state in its equilibrium configuration is shown at the bottom. Here $N=100,l=50,r=0.1$ and the distribution is obtained after 10000 iterations starting from a uniform distribution of phases.}
\end{center}
\end{figure}

\newpage
\begin{figure}[h]
\begin{center}
\begin{tabular}{c}
        \resizebox{10cm}{!}{\includegraphics[angle=0]{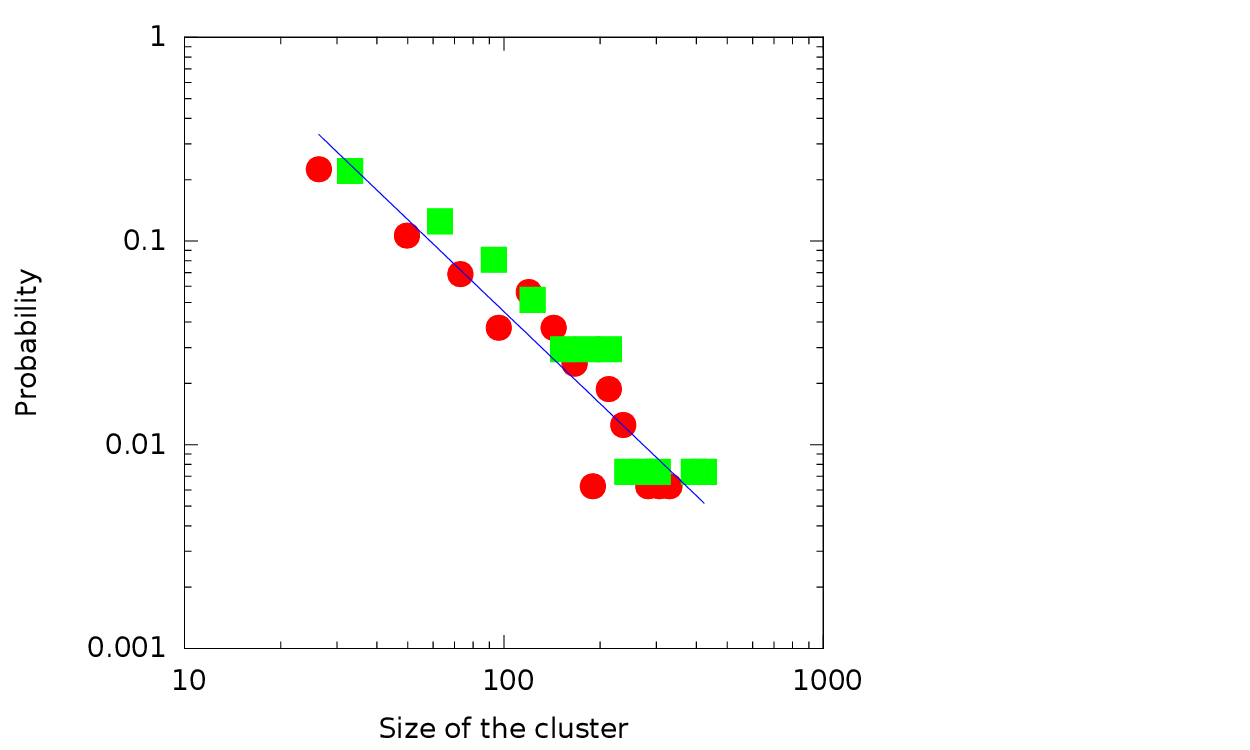}}
        
          \end{tabular}
 \caption{Cluster size Distribution of intermediate ordered states for $\epsilon=1.0,r=0.4$ (bullets) and $\epsilon=0.9,r=0.42$ (boxes). The fitted straight line is the function $45x^{-1.5}$ and we have taken $N=10000$.}
\end{center}
\end{figure}
 
\end{document}